\documentclass{aa}
\usepackage[varg]{txfonts}
\usepackage{graphicx}
\usepackage[version=3]{mhchem}
\bibpunct{(}{)}{;}{a}{}{,} 

\usepackage{color}

\begin{document}

\title{A low-mass protostar's disk-envelope interface: disk-shadowing evidence from ALMA \ce{DCO+} observations of VLA1623}

\author{N. M. Murillo\inst{1,2}
\and S. Bruderer\inst{1}
\and E. F. van Dishoeck\inst{1,3}
\and C. Walsh\inst{3}
\and D. Harsono\inst{3, 4}\thanks{Current address: Universit\"{a}t Heidelberg, Zentrum f\"{u}r Astronomie, Institut f\"{u}r Theoretische Astrophysik, Albert-Ueberle-Str. 2, 69120 Heidelberg, Germany}
\and S.-P. Lai\inst{2, 5}
\and C. M. Fuchs\inst{6}}

\institute{Max-Planck-Insitut f\"{u}r extraterrestrische Physik, Giessenbachstra\ss e 1, 85748, Garching bei M\"{u}nchen, Germany
\and Institute of Astronomy and Department of Physics, National Tsing Hua University, 101 Section 2 Kuang Fu Road, Hsinchu 30013, Taiwan
\and Leiden Observatory, Leiden University, P.O. Box 9513, 2300 RA, Leiden, the Netherlands
\and SRON Netherlands Institute for Space Research, PO Box 800, 9700 AV, Groningen, The Netherlands
\and Academia Sinica Institute of Astronomy and Astrophysics, P.O. Box 23-141, Taipei 10617, Taiwan
\and Institute of Astronautics, Technical University Munich, Boltzmannstra\ss e 15, 85748 Garching bei M\"{u}nchen, Germany}

\date{}

\abstract
{{Historically, due to instrumental limitations and a lack of disk detections, the structure of the transition from the envelope to the rotationally supported disk has been poorly studied. This is now possible with ALMA through observations of CO isotopologues and tracers of freezeout.  Class 0 sources are ideal for such studies given their almost intact envelope and young disk.}}
{{The structure of the disk-envelope interface of the prototypical Class 0 source, VLA1623A, which has a confirmed Keplerian disk, is constrained through modeling and analysis of ALMA observations of \ce{DCO+} (3-2) and \ce{C^{18}O} (2-1) rotational lines.}}
{{The physical structure of VLA1623 is obtained from the large-scale SED and dust continuum radiative transfer modeling.  An analytic model using a simple network coupled with radial density and temperature profiles is used as input for a 2D line radiative transfer calculation for comparison with the ALMA Cycle 0 12-m array and Cycle 2 ACA observations of VLA1623.}}
{{The \ce{DCO+} emission shows a clumpy structure bordering VLA1623A’s Keplerian disk. This suggests a cold ring-like structure at the disk-envelope interface. The radial position of the observed \ce{DCO+} peak is reproduced in our model only if the region’s temperature is between 11 K and 16 K, lower than expected from models constrained by continuum data and source SED. Altering the density profile has little effect on the \ce{DCO+} peak position, but increased density is needed to reproduce the observed \ce{C^{18}O} tracing the disk.}}
{{The observed \ce{DCO+} (3-2) emission around VLA1623A is the product of shadowing of the envelope by the disk observed in \ce{C^{18}O}. Disk-shadowing causes a drop in the gas temperature outside of the disk on $>$200 AU scales, encouraging the production of deuterated molecules. This indicates that the physical structure of the disk-envelope interface differs from the rest of the envelope, highlighting the drastic impact that the disk has on the envelope and temperature structure. The results presented here show that \ce{DCO+} is an excellent cold temperature tracer.}}


\titlerunning{The physical and temperature structure of the disk-envelope interface}
\authorrunning{N. Murillo et al.}

\maketitle

\begin{figure*}
\centering
\includegraphics[trim=26mm 40mm 36mm 70mm,clip,angle=-90,width=\textwidth]{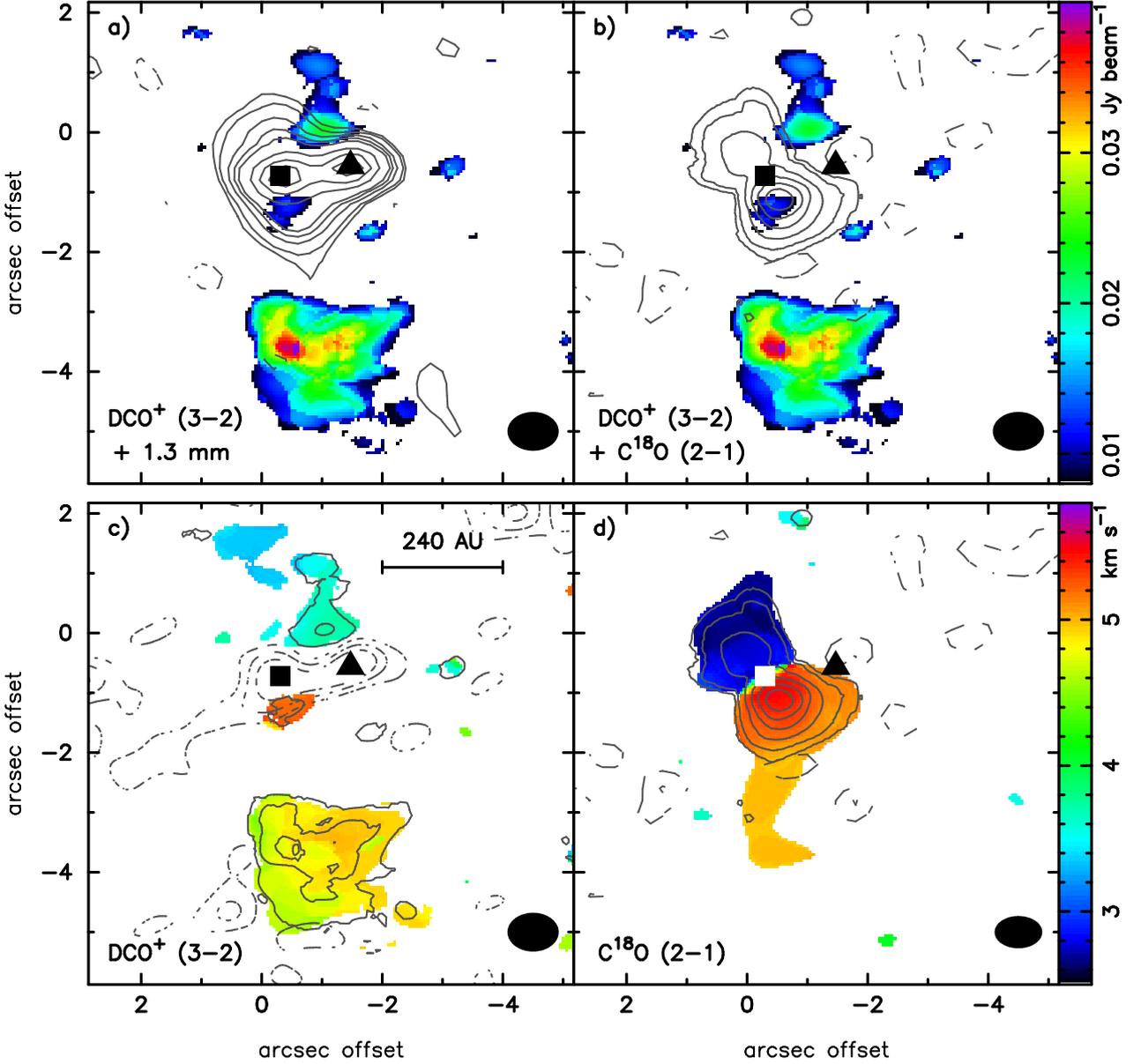}
\caption{\ce{DCO+} compared with continuum and \ce{C^{18}O} toward VLA1623A \& B, marked with a square and triangle, respectively. The black ellipse shows the synthesized beam. Intensity integrated \ce{DCO+} (color-scale) with (\textit{a}) 1.3mm continuum and (\textit{b}) \ce{C^{18}O}. \ce{DCO+} peaks at 0.0433 Jy beam$^{-1}$ km~s$^{-1}$. Contours are in steps of 3, 5, 10, 15, 20, 40, 60 and 78$\sigma$ with $\sigma$ = 1~mJy~beam$^{-1}$ for 1.3~mm continuum, and -5, -3, 3, 5, 10, 15, 20 and 25$\sigma$ with $\sigma$ = 13~mJy~beam$^{-1}$ km~s$^{-1}$ for \ce{C^{18}O}. Velocity map (moment 1, color-scale) for (\textit{c}) \ce{DCO+} and (\textit{d}) \ce{C^{18}O} overlaid with the corresponding intensity integrated map (contours). Contours are in steps of -10, -7, -4, 4, 7, 10 and 11$\sigma$ with $\sigma$ = 3~mJy~beam$^{-1}$ km~s$^{-1}$ for \ce{DCO+}, and the same as in (\textit{b}) for \ce{C^{18}O}.}
\label{figmom}
\end{figure*}

\begin{figure}
\includegraphics[trim=15mm 30mm 30mm 80mm,clip,angle=-90,width=\columnwidth]{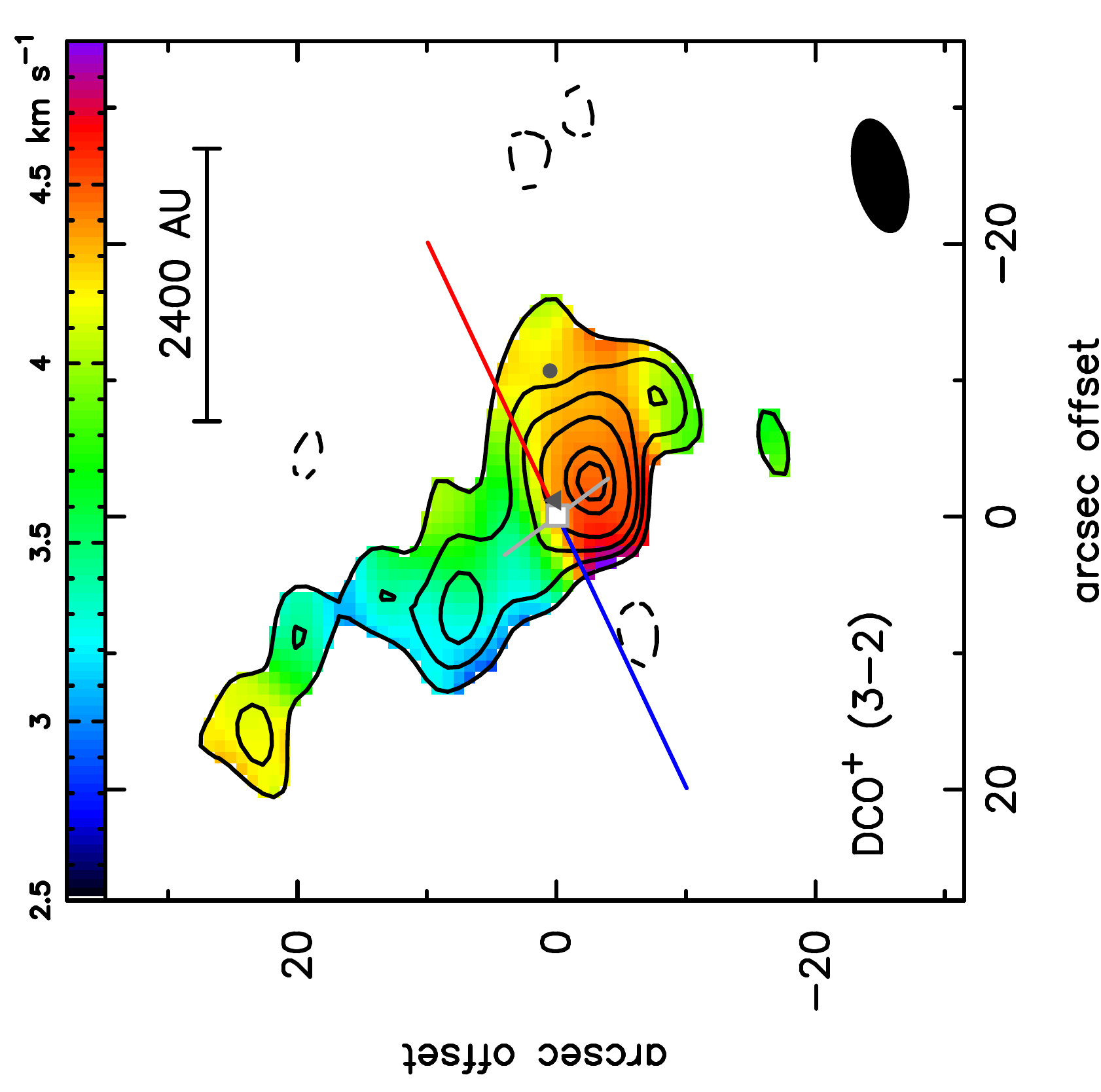}
\caption{Intensity (contours) and velocity (color-scale) integrated maps of \ce{DCO+} ACA observations. Note the 10 times larger scale of this figure compared with Fig.~\ref{figmom}. The positions of VLA1623A, B and W are marked with a square, triangle and circle, respectively. The size of the filled square shows the size of the \ce{C^{18}O} disk. The red and blue lines indicate the outflow direction while the gray line shows the mayor axis of the disk plane. The black ellipse shows the synthesized beam. Contours are in steps of -3, 3, 5, 7, 10, 15 and 17$\sigma$ with $\sigma$ = 220~mJy beam$^{-1}$ km~s$^{-1}$.}
\label{figacamom}
\end{figure}

\section{Introduction}
\label{secintro}
Rotationally supported disks have been observed extensively among most protostellar and pre-main sequence evolutionary stages \citep{li2014}. Recent studies have revealed the existence of such disks in the Class 0 deeply embedded phase \citep{choi2010, tobin2013, murillo2013b, codella2014, lee2014}. In the early stages of star formation, the envelope is not yet dispersed and contains enough material to influence the evolution of the star-disk system. The boundary between the disk and the envelope, known as the disk-envelope interface, must then play a role in the formation process. This region, however, is largely unexplored owing to limitations in the resolution and sensitivity, as well as the lack of observed rotationally supported disks, until now. Class 0 sources with confirmed rotationally supported disks grant us the opportunity to study the chemical and physical structure of the disk-envelope interface region, which is crucial for the next step of understanding this region's role in star formation \citep{sakai2014}.

Whilst \ce{CO} isotopologues are good tracers of rotationally supported disks, they freeze out onto dust grains below the evaporation temperature $T_{\rm ev}$ usually between 30-20 K \citep{jorgensen2005}. These low temperatures are reached at the edge of the embedded disk \citep{visser2009}. As a result molecular species whose abundance is enhanced at low temperatures are needed to trace the disk-envelope interface. \ce{DCO+} emission is known to be optically thin and its abundance is enhanced at a narrow range of temperatures below the \ce{CO} freeze-out temperature tracing the so-called \ce{CO} snowline (e.g., \citealt{wootten1987}, \citealt{roberts2003}, \citealt{mathews2013}). Thus, \ce{DCO+} is a good candidate molecule to trace the chemical, physical and kinematic structure of the disk-envelope interface.

VLA1623-2417 (hereafter VLA1623) is a triple non-coeval protostellar system located in $\rho$ Ophiuchus at $d\sim$120 pc \citep{murillo2013a}. VLA1623A is the prototypical Class 0 source and emits predominantly in the (sub)millimeter range \citep{andre1993, murillo2013a}. From modeling of ALMA Cycle 0 \ce{C^{18}O} observations, \cite{murillo2013b} found that VLA1623A supports a Keplerian disk with a radius of at least 150 AU and a central mass $M_{\rm *}$ of 0.2 M$_{\odot}$. The ALMA Cycle 0 observations also detected \ce{DCO+} molecular line emission toward VLA1623A bordering the \ce{C^{18}O} disk. This grants us the opportunity to probe the disk-envelope interface of VLA1623A.

In this paper we present the results of our ALMA observations and simple chemical modeling, aiming to understand the physical structure of the boundary between the envelope and the disk in a Class 0 protostar.

\begin{figure}
\includegraphics[trim=50mm 80mm 40mm 75mm,clip,angle=-90,width=\columnwidth]{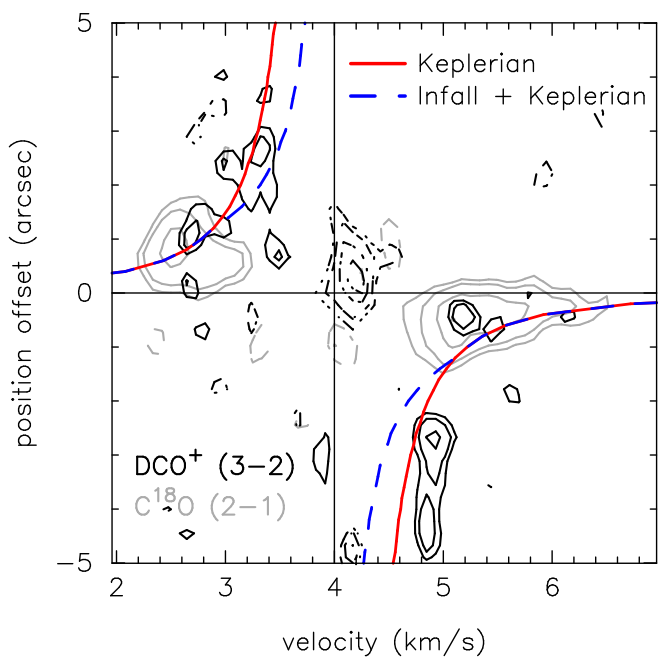}
\caption{Cycle 0 \ce{DCO+} (black) and \ce{C^{18}O} (gray) PV diagram. Cuts are made at P.A. = 26$^{\circ}$ for \ce{DCO+} and P.A. = 35$^{\circ}$ for \ce{C^{18}O}. Keplerian (solid line) and Infall plus Keplerian (dashed line) rotation PV models with $M_{*}$ = 0.2 M$_{\odot}$ are overplotted \citep{murillo2013b}. Contours are in steps of -3, 3, 5, 10 and 15$\sigma$ with $\sigma$ = 19~mJy~beam$^{-1}$ for \ce{C^{18}O}, -7, -5, -3, -2, 2, 3, 4, 5$\sigma$ with $\sigma$ = 12~mJy~beam$^{-1}$ for Cycle 0 \ce{DCO+}}
\label{figpv} 
\end{figure}

\section{Observations}
\label{secobs}
Using the Atacama Large Millimeter/submillimeter Array (ALMA) we observed VLA1623 (pointing coordinates $\alpha$=16:26:26.419 $\delta$=-24:24:29.988 J2000) during the early science Cycle 0 period on 8 April 2012. 
Observations were carried out using the extended configuration, comprised of 16 antennas with a maximum baseline of $\sim$400~m, in Band 6 (230 GHz). 
Total observing time was one hour with a 58\% duty cycle. 
The spectral configuration was set to observe \ce{DCO+} (3-2), \ce{C^{18}O} (2-1), and \ce{^{12}CO} (2-1) in addition to continuum. 
\ce{N2D+} was also observed but no significant detection was made. 
Data calibration was done with 3C 279, 1733-130, and Titan for bandpass, gain, and flux calibration, respectively. 

In this paper, we present the results and analysis of the \ce{DCO+} (3-2) (rest frequency: 216.11258 GHz) observations. The \ce{DCO+} data were calibrated jointly with the continuum, \ce{C^{18}O} (2-1), and \ce{^{12}CO} (2-1) data. Further calibration details and results from the other observed lines can be found in \cite{murillo2013b}. The spectral set-up provided a velocity resolution of 0.0847 km s$^{-1}$. The synthesized beam size for the \ce{DCO+} images is 0.85$\arcsec$ $\times$ 0.65$\arcsec$ with P.A. = 96$^{\circ}$. The rms noise of the channel map is 12 mJy beam$^{-1}$ for a spectral resolution of 0.0847 km s$^{-1}$, giving a peak S/N = 7. 

Our ALMA observations provide a maximum scale of 4$\arcsec$ and a field of view (FOV) of 24$\arcsec$, with emission between 4$\arcsec$ and 24$\arcsec$ largely filtered out by the interferometer. The FOV together with the beam size of 0.85$\arcsec$ constrain the scale to which any analysis of the data can be done. 

In addition to the ALMA Cycle 0 observations, we present the \ce{DCO+} (3-2) results from our ALMA Cycle 2 Atacama Compact Array (ACA) observations carried out on 7 August 2014 (pointing coordinates $\alpha$=16:26:26.390 $\delta$={-24:24:30.688}). Total observing time was 2 hours. Data calibration was done with J1517-243 and Mars for flux, J1625-2527 and Mars for gain, and J1733-1304 for bandpass. The rms noise of the \ce{DCO+} channel map is 73 mJy~beam$^{-1}$ for a spectral resolution of 0.021 km~s$^{-1}$, with a synthesized beam of 8.6$\arcsec \times 4.2\arcsec$ with P.A. = -76$^{\circ}$. These observations, with a mosaicked area of 6$\arcmin$, provide the \ce{DCO+} emission between 4$\arcsec$ and 18$\arcsec$ scales.

\section{Results}
\label{secres}
\subsection{ALMA 12-m Array Cycle 0}
The detected \ce{DCO+} emission (Fig.~\ref{figmom}) is located between a velocity range of 2.8 to 5.2 km~s$^{-1}$ and shows a clumpy structure with two main clumps to the north and south of VLA1623A. The southern clump emission is stronger and is offset by about 2.5$\arcsec$ (300 AU) from VLA1623A and in addition it borders the red-shifted emission of the disk traced in \ce{C^{18}O}. The northern clump slightly overlaps the blue-shifted emission of the \ce{C^{18}O} disk and borders VLA1623B's continuum emission. A couple of clumps with emission between 3 and 10$\sigma$ are observed near the continuum peaks of VLA1623A \& B. No significant emission was detected toward VLA1623W, separated by 10$\arcsec$ to the west from VLA1623A, possibly either because of the lack of \ce{DCO+} or because the emission is too weak and filtered out. 

The \ce{C^{18}O} line emission tracing the disk was found to be influenced by the outer envelope, with the blue-shifted emission being affected more than the red-shifted emission \citep{murillo2013b}. We expect the same to hold true for the \ce{DCO+} emission, thus potentially explaining why the northern blue-shifted clump emission is weaker than that for the southern red-shifted clump (Fig.~\ref{figmom} a \& b). 

The velocity weighted (moment 1) map of the \ce{DCO+} emission (Fig.~\ref{figmom}c) shows that its velocity gradient is similar to that of \ce{C^{18}O}, with the northern clump being blue-shifted and the southern clump being red-shifted, but with a smaller velocity range. The Position-Velocity (PV) diagrams of \ce{DCO+} and \ce{C^{18}O} emission (Fig.~\ref{figpv}) are constructed and over-plotted with the best fitting thin disk models, Keplerian and Infall plus Keplerian out to 150 AU, obtained by \cite{murillo2013b}. It appears as though both line emissions are well described by pure Keplerian rotation out to 300 AU. However the \ce{DCO+} emission is too weak, peaking at 7$\sigma$ in the channel map, to carry out further kinematical analysis. The similar velocity signatures of the emission from both species may be due to the disk edge dragging along material from the envelope.

The above simple analysis of the observations indicates that the detected \ce{DCO+} emission may be tracing a ring which borders the \ce{C^{18}O} disk. In addition, \ce{DCO+}'s velocity gradient and PV diagram suggest that the ring is undergoing Keplerian rotation, this is even more plausible given that the \ce{C^{18}O} (2-1) emission traces a rotationally supported disk with a Keplerian velocity profile \citep{murillo2013b}.

\subsection{ALMA ACA Cycle 2}
The \ce{DCO+} (3-2) emission traced with the ACA confirms that the emission is concentrated around VLA1623 (Fig.~\ref{figacamom}), it does not peak on VLA1623A and instead encircles it, forming a shell-like structure. The ACA detects only weak emission at the position of VLA1623W.

The \ce{DCO+} emission mapped with the ACA shows the same velocity gradient (Fig.~\ref{figacamom}) as on the small scales, indicating that the kinematic structure is the same throughout.

\begin{figure}
\centering
\includegraphics[scale=0.7]{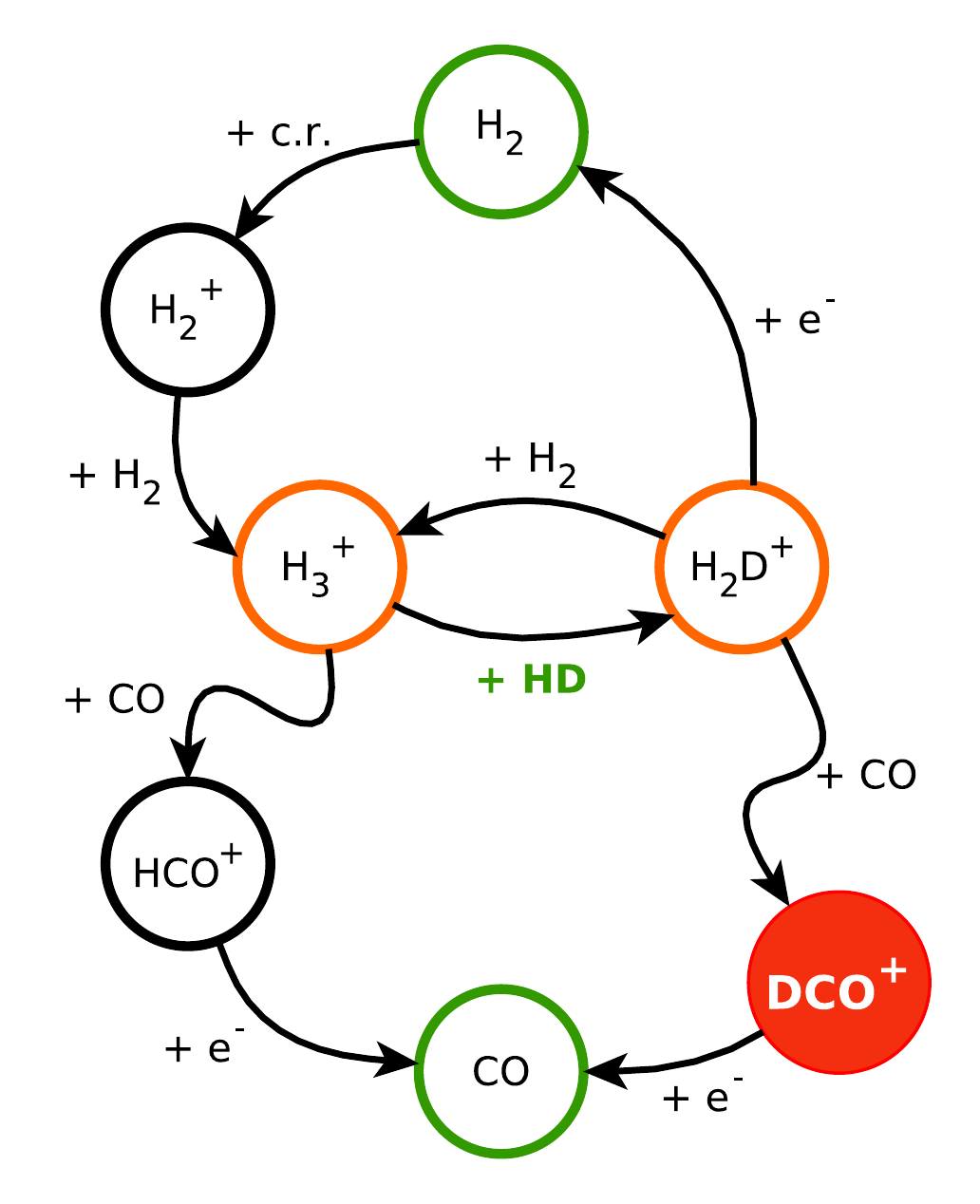}
\caption{Cartoon of the chemical network used to analyze the observed \ce{DCO+} emission. Red denotes the target molecule. Green the input molecular concentrations and orange the species involved in the bottleneck reaction.}
\label{fignet}
\end{figure}

\begin{table*}
\caption{\ce{DCO+} chemical network reactions and adopted rate coefficients}
\label{tabnet}
\centering
\begin{tabular}{l l c c c c c}
\hline \hline
ID & Reaction & $\zeta$\tablefootmark{a} & $\alpha$\tablefootmark{b} & $\beta$\tablefootmark{c} & $\gamma$\tablefootmark{d} & Ref.\\
 & & s$^{-1}$ & cm$^{3}$ s$^{-1}$ & & K & \\
\hline
1 & \ce{H2 + cr -> H2+ + e-} & 1.26 (-17) & ... & ... & ... & 1 \\
2 & \ce{H2+ + H2 -> H3+ + H} & ... & 2.08 (-9) & 0 & 0 & 2 \\
3 & \ce{H3+ + HD -> H2D+ + H2} & ... & 3.50 (-10) & 0 & 0 & 3, 4 \\
4 & \ce{H2D+ + H2 -> H3+ + HD} & ... & 3.50 (-9) & 0 & 220.0 & 3, 4 \\
4a & \textbf{\ce{H2D+ + p-H2 -> H3+ + HD}} & ... & 1.40 (-10) & 0 & 232.0 & 5 \\
4b & \textbf{\ce{H2D+ + o-H2 -> H3+ + HD}} & ... & 7.00 (-11) & 0 & 61.5 & 5 \\
5 & \ce{H2D+ + e- -> HD + H} & ... & 6.79 (-8) & -0.52 & 0 & 5\tablefootmark{e} \\
6 & \ce{H3+ + CO -> HCO+ + H2} & ... & 1.61 (-9) & 0 & 0 & 6, 7 \\
7 & \ce{HCO+ + e- -> H + CO} & ... & 2.80 (-7) & -0.69 & 0 & 8 \\
8 & \ce{H2D+ + CO -> DCO+ + H2} & ... & 5.37 (-10) & 0 & 0 & 4 \\
9 & \ce{DCO+ + e- -> D + CO} & ... & 2.40 (-7) & -0.69 & 0 & 4 \\
\hline
\end{tabular}
\\
\tablefoot{The reactions in bold (4a and 4b) substitute the back reaction of the bottleneck (4) in the network when \ce{o-H2} and \ce{p-H2} are included.\\
\tablefoottext{a}{$\zeta$ is the cosmic-ray ionization rate of \ce{H2}.}
\tablefoottext{b}{$\alpha$ is the rate coefficient at $T$ = 300 K.} 
\tablefoottext{c}{$\beta$ is the temperature coefficient. } 
\tablefoottext{d}{$\gamma$ is the activation barrier.}
\tablefoottext{e}{Value obtained by summing the rate coefficients for the reaction channels to three possible products.}
}
\tablebib{(1)~\citealt{black1975}; (2)~\citealt{theard1974}; (3)~\citealt{gerlich2002}; (4)~\citealt{albertsson2013}; (5)~\citealt{walmsley2004}; (6)~\citealt{plessis2010}; (7)~OSU Database; (8)~\citealt{amano1990}}
\end{table*}

\section{Analysis}
\label{secanal}
\ce{DCO+} is detected at the disk-envelope interface of VLA1623A. Given that \ce{DCO+} is optically thin and a good probe of temperature and CO freeze-out regions, we model the observed emission aiming to probe the physical and chemical structure of the disk-envelope interface. In this section we describe the model used and the results obtained from the model.

\subsection{\ce{DCO+} chemical network and model}
\label{subsecchemnet}
We model the observed \ce{DCO+} emission using a simple chemical network. Such a network, while it may not account for every possible reaction, is preferred as it gives insight into how the physical parameters of the temperature and density affect the observed emission. The network used is a steady-state analytic model that only takes the basic reactions that lead to \ce{DCO+} production and destruction into account (Table~\ref{tabnet}, Fig.~\ref{fignet}). Since \ce{DCO+} is formed by the reaction of \ce{H2D+} and \ce{CO}, the rate-determining reactions in our network are
\begin{equation}
\label{eqforward}
\cee{H3+ + HD -> H2D+ + H2}
\end{equation}
with the back reaction
\begin{equation}
\label{eqbn}
\cee{H2D+ + H2 + $\Delta$E -> H3+ + HD}
\end{equation}
where the activation energy due to the difference in zero point energy is $\Delta$E $\sim$ 220~K. For reaction 5 in our network (Table~\ref{tabnet}), for simplicity, we adopt the total rate coefficient summing over all three pathways of the reaction. The rate coefficients for a two-body reaction are given by
\begin{equation}
\label{eqratecoeff2b}
k = \alpha \left(\frac{T}{300}\right)^\beta {\rm exp}\left(-\frac{\gamma}{T}\right) ~\rm cm^{3}~s^{-1}
\end{equation}
where T is the gas temperature, while
\begin{equation}
\label{eqratecoeff1b}
k = \zeta ~\rm s^{-1}
\end{equation}
gives the rate coefficient for cosmic-ray ionization (reaction 1 from Table~\ref{tabnet}).

Our model takes as input a source density and temperature profile as a function of radius and the parameters needed to calculate the rate coefficients. The \ce{CO} evaporation temperature $T_{\rm ev}$, desorption density $n_{\rm de}$ and CO abundance $X_{\rm CO}$ are free parameters. The \ce{CO} abundance is assumed and not calculated. The \ce{CO} evaporation temperature dictates when \ce{CO} is in the gas phase ($T$ > $T_{\rm ev}$) or freezes onto the dust grains ($T$ < $T_{\rm ev}$). In a similar manner, the desorption density sets the boundary when the freeze-out timescales are too long ($n$ < $n_{\rm de}$) compared to the lifetime of the core \citep{jorgensen2005}. We assume the density profile is equal to the \ce{H2} density $n_{\rm \ce{H2}}$ and the abundance of \ce{HD} $X_{\rm \ce{HD}}$ = 10$^{-5}$ with respect to the total hydrogen nuclei density $n_{H}$ = $2n$(\ce{H2}). The model returns the calculated concentrations as a function of radius. The results can then be input into excitation radiative transfer programs such as RATRAN \citep{hogerheijde2000} for further analysis.

\begin{figure}
\includegraphics[width=\columnwidth]{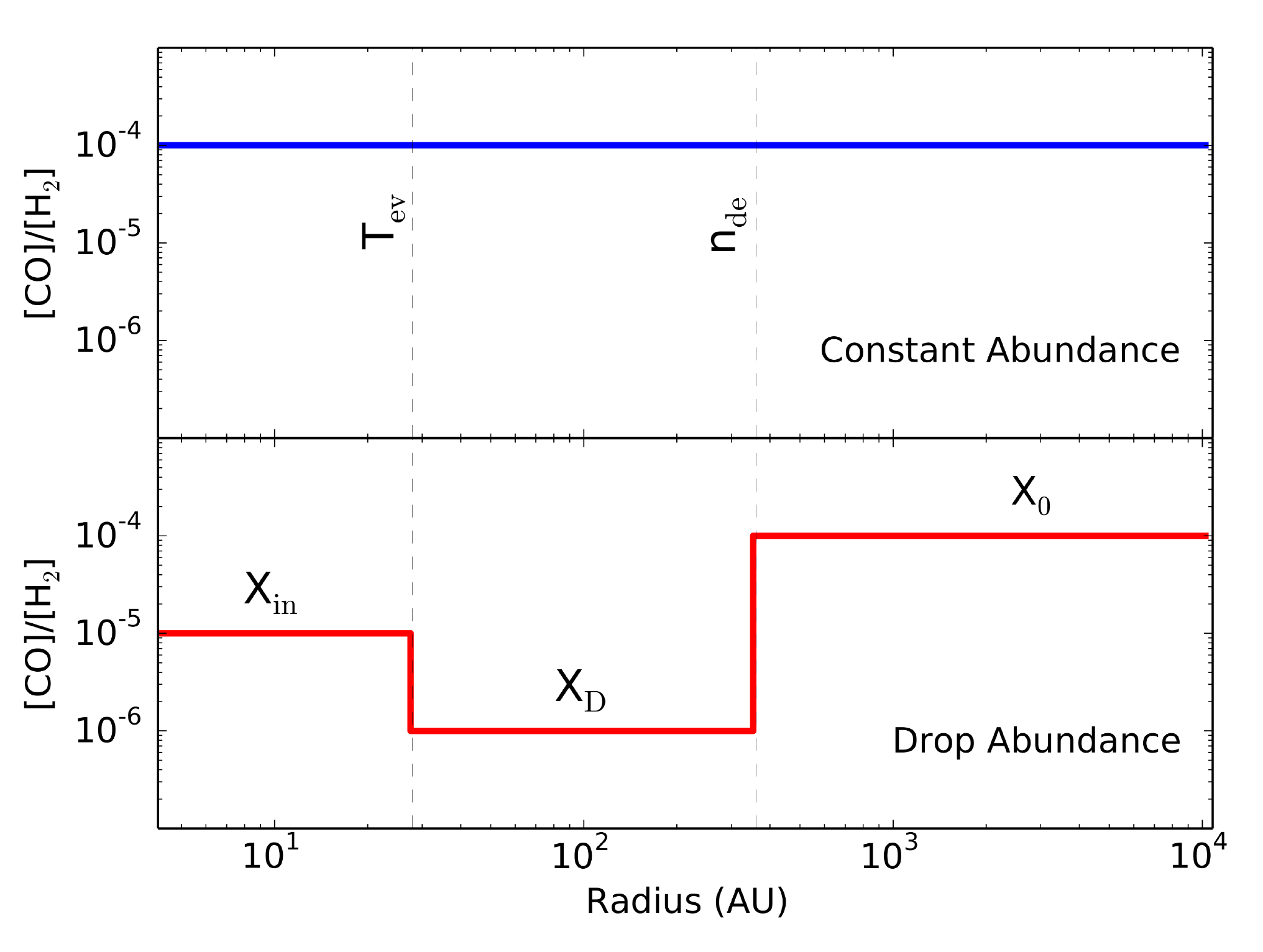}
\caption{CO abundance profiles used in the model. The vertical dashed lines show the limits for the Drop abundance profile, evaporation temperature $ T_{\rm ev} $ and desorption density $ n_{de} $. $X_{\rm 0}$ denotes the abundance in the constant profile. $X_{\rm in}$, $X_{\rm D}$ and $X_{\rm 0}$ are the inner, drop and outer region abundances for the drop profile.}
\label{figpro}
\end{figure}

Given that \ce{CO} is one of the parent molecules of \ce{DCO+}, we study the effect of \ce{CO} abundance through the use of different abundance profiles following the models detailed in \cite{jorgensen2005} and \cite{yildiz2010}. The possible \ce{CO} profiles are Constant and Drop abundance (Fig.~\ref{figpro}). The former represents a fixed \ce{CO} abundance throughout the core, whereas the latter, constrained by $T_{\rm ev}$ and $n_{de}$, is used to account for \ce{CO} freeze-out in the chemical network. For the constant profile, the abundance is denoted by $X_{\rm 0}$. For the drop profile, the abundance in the inner, drop and outer regions are denoted by $X_{\rm in}$, $X_{\rm D}$ and $X_{\rm 0}$, respectively. 
Previous studies using multiline single dish \ce{C^{18}O} have found that the abundance at $X_{\rm in}$ is lower than $X_{\rm 0}$ for a number of sources \citep{alonso2010, yildiz2010, yildiz2013}. One explanation is that some fraction of the CO ice is transformed into more complex and less volatile carbonaceous species in the cold phase. We thus take this effect in our model into account.
A lower abundance of \ce{CO}, due to freeze-out, allows an increase in the abundance of \ce{H2D+} \citep{mathews2013}. However, since both molecules, \ce{CO} and \ce{H2D+}, are parent molecules of \ce{DCO+}, a balance must be reached before the effective formation of \ce{DCO+} takes place. This scenario is found to be common for the envelopes of early embedded protostars \citep{jorgensen2005} where the outer region shielded from the protostellar and interstellar radiation heating has low enough temperature that \ce{CO} freezes out onto dust grains.

In molecular clouds, and consequently in protostellar cores, \ce{H2} chemistry plays a major role, hence the ortho-to-para ratio of \ce{H2} influences the chemical reactions, and has been found to be crucial to the deuterium chemistry \citep{flower2006, pagani2009}. The effect of ortho- and para-\ce{H2} (o-\ce{H2} and p-\ce{H2}) is studied in our model through the inclusion of the ortho-to-para ratio (o/p) and the distinction of o-\ce{H2} and p-\ce{H2} in the back reaction of the bottleneck (Eq.~\ref{eqbn}) in the chemical network. o-\ce{H2} and p-\ce{H2} are only added in the back reaction since it is here where the distinction has a significant effect (Table~\ref{tabnet}). We set a lower limit on o/p of 10$^{-3}$ at low temperatures, as constrained by models and observations \citep{flower2006,faure2013}. The o-\ce{H2} and p-\ce{H2} reactions and their parameters for the rate coefficients are taken from \cite{walmsley2004}.

When o-\ce{H2} and p-\ce{H2} are included in the network, a thermal (LTE), upper- or lower-limit o/p ratio can be selected. In LTE, the ortho-to-para ratio is given by
\begin{equation}
{\rm o/p} = 9 \exp \left( \frac{-170}{T} \right)
\label{eqop}
\end{equation}
where $T$ is the gas temperature. Selecting the upper-limit ratio produces 3 times more o-\ce{H2} than p-\ce{H2}. Since the back reaction with o-\ce{H2} has a lower activation barrier $\gamma$ than with p-\ce{H2}, the o/p upper-limit implies that \ce{H_2D+} is being destroyed faster than generated, leading to a decreased production of \ce{DCO+} (Fig.~\ref{figdropoplim}), since \ce{H_2D+} is a parent molecule of \ce{DCO+} (Fig.~\ref{fignet}). The lower-limit ratio, on the other hand, implies more p-\ce{H2} which has a higher activation barrier $\gamma$ for the back reaction, thus \ce{H_2D+} is generated faster than it is destroyed in turn increasing the \ce{DCO+} production (Fig.~\ref{figdropoplim}).

As a starting point for our analysis of VLA1623A's \ce{DCO+} emission, we use the density and temperature profile of VLA1623 obtained by \cite{jorgensen2002} where 30 K is at $\sim$1.5" assuming a distance of 120 pc. The profile was obtained by fitting single dish JCMT continuum images and the spectral energy distribution (SED) with continuum radiative transfer modeling using DUSTY resulting in a power law density profile of the form $n \propto r^{-1.4}$. In the single dish continuum observations VLA1623A and B are unresolved and the density profile extends well beyond the ALMA field of view. It is expected, however, that VLA1623A dominates at 870 $\mu$m and that VLA1623B does not contribute much to the 450 $\mu$m continuum \citep{murillo2013a}. Thus, the density and temperature profile obtained by \cite{jorgensen2002} is representative of VLA1623A since VLA1623B is not significantly contributing to the continuum emission or the SED used to constrain the profiles.

The results of the analytic chemical network are run through the molecular excitation and line radiative transfer program RATRAN to generate line emission maps. Since the structure we are trying to reproduce is ring-like, we calculate the level populations with the 1-D version of RATRAN, and then run the level populations with the 2-D ray tracing to form the ring structure. The produced spectral image cubes are convolved with a Gaussian beam with the dimensions of the synthesized beam, continuum subtracted and then an intensity integrated map is generated. Radial profiles are extracted from the resulting images and compared with the observed profiles, which are integrated over the extent of the detected emission. 

We find no significant difference in the peak position between the method used here and running the models through the ALMA simulator with the actual ALMA configuration or our Cycle 0 observations. However, the ALMA simulator does show that the emission $\geq$4$\arcsec$ is indeed filtered out by the Cycle 0 observations, producing somewhat narrower radial profiles.
Molecular data for \ce{DCO+} and \ce{C^{18}O} are obtained from LAMDA (\citealt{schoeier2005}; \ce{C^{18}O}: \citealt{yang2010}; \ce{DCO+} extrapolated from \citealt{flower1999}). As \ce{DCO+} is optically thin, the comparison mainly focuses on the position of the peak and the integrated intensity profile with respect to radius, the velocity profile has no effect on the resulting model integrated intensity. Thus, we assume a free-fall velocity profile with a central mass of 0.2 $M_{\rm \odot}$ \citep{murillo2013b} and $T_{\rm dust}$ = $T_{\rm gas}$. The RATRAN output maps are convolved with the observed beam (0.85$\arcsec$ $\times$ 0.65$\arcsec$, P.A. 96.24$^{\circ}$) and compared with the observed emission through radial cuts. 
In addition to comparing \ce{DCO+}, we also compare \ce{C^{18}O} in order to further constrain the physical structure of the disk-envelope interface. However, for \ce{C^{18}O} we assume a pure Keplerian velocity profile, in agreement with the rotationally supported disk it traces. For both emission lines, we use an inclination of 55$^{\circ}$ (90$^{\circ}$ = face-on), in accordance with the results obtained by \cite{murillo2013b}.

\begin{table}
\caption{Input CO abundance profiles}
\label{tabcase}
\begin{tabular}{c c c c c c c}
\hline \hline
Case & \ce{CO} & $T_{\rm ev}$ & $n_{\rm de}$ & $X_{\rm in}$ & $X_{\rm D}$ & $X_{\rm 0}$ \\
 & profile & (K) & (cm$ ^{-3} $) & & & \\
\hline 
Fixed & Constant & ... & ... & ... & ... & 10$^{-4}$ \\
a & Drop & 30 & 7$\times$10$^{4}$ & 10$^{-5}$ & 10$^{-6}$ & 10$^{-4}$ \\
b & Drop & 35 & 3$\times$10$^{6}$ & 10$^{-5}$ & 10$^{-6}$ & 10$^{-4}$ \\
c & Drop & 35 & 3$\times$10$^{6}$ & 10$^{-5}$ & 5$\times$10$^{-6}$ & 10$^{-4}$ \\
d & Drop & 35 & 3$\times$10$^{6}$ & 10$^{-5}$ & 10$^{-7}$ & 10$^{-4}$ \\
e & Drop & 35 & 5$\times$10$^{5}$ & 10$^{-5}$ & 10$^{-6}$ & 10$^{-4}$ \\
f & Drop & 35 & 3$\times$10$^{6}$ & 10$^{-5}$ & 10$^{-8}$ & 10$^{-4}$ \\
\hline
\end{tabular}
\tablefoot{See Fig.~\ref{figpro} for definition of $X_{\rm in}$, $X_{\rm D}$ and $X_{\rm 0}$}
\end{table}

\begin{figure*}
\includegraphics[width=\textwidth]{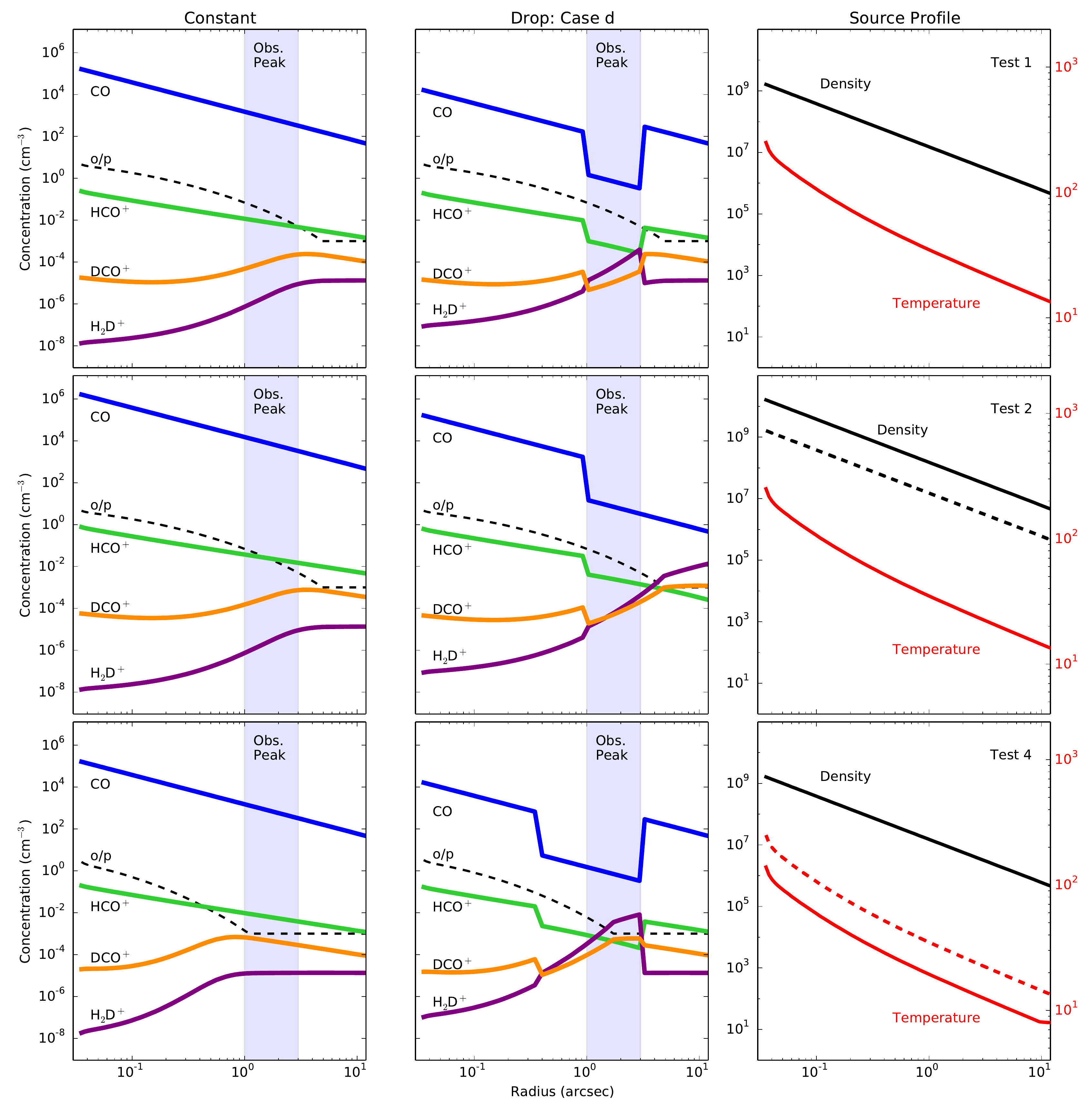}
\caption{Chemical network model results. Left and center columns show the results for constant and drop CO abundance, respectively. Top, middle and bottom row show the results of Test 1, 2 and 4, respectively. Light blue shaded region shows the observed peak position of \ce{DCO+}. Right column shows the input source profile, where the original DUSTY source profile is shown with a dashed line so as to compare with the input for the specific test.}
\label{figresults}
\end{figure*}

\begin{table*}
\caption{Input source profiles for each test}
\label{tabinput}
\centering
\begin{tabular}{c c c c}
\hline
Test & $T_{\rm 4AU}$ (K) & $n_{\rm 4AU}$ (cm$^{-3}$) & Source Profile \\
\hline
1 & 250.0 & 1.62 $\times$ 10$^{9}$ & from \cite{jorgensen2002} \\
2 & 250.0 & 1.62 $\times$ 10$^{10}$ & Density increased 1 order of magnitude \\
3 & 250.0 & 1.62 $\times$ 10$^{8}$ & Density decreased 1 order of magnitude \\
4 & 166.7 & 1.62 $\times$ 10$^{9}$ & Temperature decreased by a factor of 1.5 with $T_{\rm lowlim}$ = 8 K \\
\hline
\end{tabular}
\end{table*}

\subsection{Modeling results}
\label{subsecmod}
In this section the results of altering the \ce{CO} abundance, density and temperature profiles in the chemical model are discussed. Figure~\ref{figresults} presents the models, with the light blue region showing the location of the observed \ce{DCO+} peak emission with respect to VLA1623A's position. Model naming follows the scheme xYz where x is the test number from Table~\ref{tabinput}, Y is either C for constant or D for drop abundance, and z is the case from Table~\ref{tabcase}.

\subsubsection{Chemical properties}
\label{subsecmodchem}
To probe the chemical conditions, we establish one case for the constant \ce{CO} abundance with $X_{\rm CO}$ = 10$^{-4}$ and six cases for the drop \ce{CO} abundance with $X_{\rm D}$ ranging from 5$\times$10$^{-6}$ to 10$^{-8}$ and varying $T_{\rm ev}$ and $n_{\rm de}$. Parameter ranges were selected based on trends found in previous work \citep{jorgensen2005,yildiz2010} and adapted to the current observations. Parameters for each case are listed in Table~\ref{tabcase}. As a zeroth order test, we compare the abundance profiles to the radial position of the observed \ce{DCO+} peak in the ALMA 12-m array data in order to investigate which chemical conditions best approximate our observations. This comparison neglects the fact that the peak emission radius also depends on the \ce{DCO+} excitation, which will be taken in Section~\ref{subsecemission} into account.

Figure~\ref{figresults} presents the model abundance profiles as functions of radius for different assumed \ce{CO} abundances and physical structures. In general, the concentrations of \ce{CO} and \ce{HCO+} drop with radius whereas those of \ce{H_{2}D+} and \ce{DCO+} increase with radius due to the lower temperature farther away from the source. 
Constant \ce{CO} abundance (Fig.~\ref{figresults}, left column) does not appear to produce a \ce{DCO+} peak within the expected region. Since the radial abundance profile of \ce{DCO+} is not altered by changing $X_{\rm CO}$, we focus instead on the drop \ce{CO} abundance profile.


While the drop \ce{CO} abundance profile cannot alone alter the position of the peak, it produces several trends interesting to note. A kink in the abundance of \ce{DCO+} forms at $T_{\rm ev}$ for all the cases examined, though in most cases it is relatively small. Altering the abundance in the drop $X_{\rm D}$ changes the shape of the peak but does not significantly alter its position nor its abundance (Fig.~\ref{figdropcase}). Decreasing the $X_{\rm D}$ below 10$^{-7}$, however, causes the abundance of the peak to drop by several orders of magnitude and become similar in magnitude to the kink at $T_{\rm ev}$, producing two peaks which are not observed (Fig,~\ref{figdropcase}, case f). Varying $T_{\rm ev}$ and $n_{\rm de}$, in order to constrain the width of the drop, generates a very narrow peak which increases and drops quickly. 

To test the effect of the o/p ratio on our chemical network, we vary the ratio from the thermalized value to the upper- and lower-limit for the drop \ce{CO} abundance profile. In all cases, we find that setting o/p to the lower-limit does not change the position of the \ce{DCO+} emission since the peak and bulk of the \ce{DCO+} concentration for the thermalized o/p is already located in the o/p=10$^{-3}$ range (Fig.~\ref{figdropoplim}). In fact, the peak of the modeled \ce{DCO+} emission starts to decrease as the o/p ratio increases. The lower-limit o/p only alters the inner regions ($<$100 AU), which the data do not constrain.  The upper-limit reduces the overall production of \ce{DCO+} and pushes the peak outward to larger radii. In conclusion, altering the \ce{CO} abundance and o/p ratio does not produce the observed results.

\subsubsection{Physical properties}
\label{subsecmodphys}
Altering the chemistry of the model does not reproduce the observed \ce{DCO+} peak position, regardless of the case used. Thus we alter the density and temperature profile in order to find the conditions necessary to reproduce the observed emission. We set up 4 tests, including the original profile from \cite{jorgensen2002} referred to as Test 1, by increasing or decreasing by a constant factor either the density or temperature profile. Parameters for each test are listed in Table~\ref{tabinput}. 

Increasing or decreasing the density profile by one order of magnitude, Tests 2 and 3, does not generate a significant radial shift of the \ce{DCO+} peak compared with the original profile (Fig.~\ref{figresults}, middle row). Altering the density, however, affects the concentration of the peak upwards or downwards with an increase or decrease of the density, respectively. 

The factor of 1.5 decrease in the temperature profile with cases \textit{b} and \textit{d} with $X_{D} \approx 10^{-6} - 10^{-7}$ shifts the modeled \ce{DCO+} peak effortlessly to the observed position (Fig.~\ref{figresults}). Decreasing the temperature profile between a factor of 1 to 2 moves the modeled \ce{DCO+} peak inward (Fig.\ref{figaptemp}). For the constant \ce{CO} abundance, any alteration of the temperature either over-estimates the modeled \ce{DCO+} abundance or produces a peak too far inward. For the drop \ce{CO} abundance, a temperature drop less than a factor of 1.5 does not move the peak inward enough, whereas a larger factor moves it too far inward. Cases $b$ and $d$ with $X_{D} \approx 10^{-6} - 10^{-7}$ shifts the modeled \ce{DCO+} peak effortlessly to the observed position (Fig.~\ref{figresults} bottom row). Changing the \ce{CO} abundance in the drop (Table~\ref{tabcase}) does not alter the location of the peak, as expected from the results presented in Sec.~\ref{subsecmodchem}. Examining the temperature profile shows that the \ce{DCO+} emission peaks at a range of 11-16 K. This is lower than the expected 20 K at $\sim$3$\arcsec$ inferred from radiative transfer modeling of the observed continuum data \citep{jorgensen2002}. We limit the decrease in the temperature profile to not fall below 8 K \citep{zucconi2001}, however this limit generates no significant change in the outcome of our model since the limit falls near and beyond the edge of the FOV of our observations (Fig.~\ref{figresults}, bottom right panel). 

From the results obtained from altering the chemical and physical conditions in our model, we can deduce that the \ce{DCO+} emission is located at its observed position due to physical conditions, namely a lowered temperature, and not to special chemical conditions in the disk-envelope interface.

\begin{figure}
\includegraphics[width=\columnwidth]{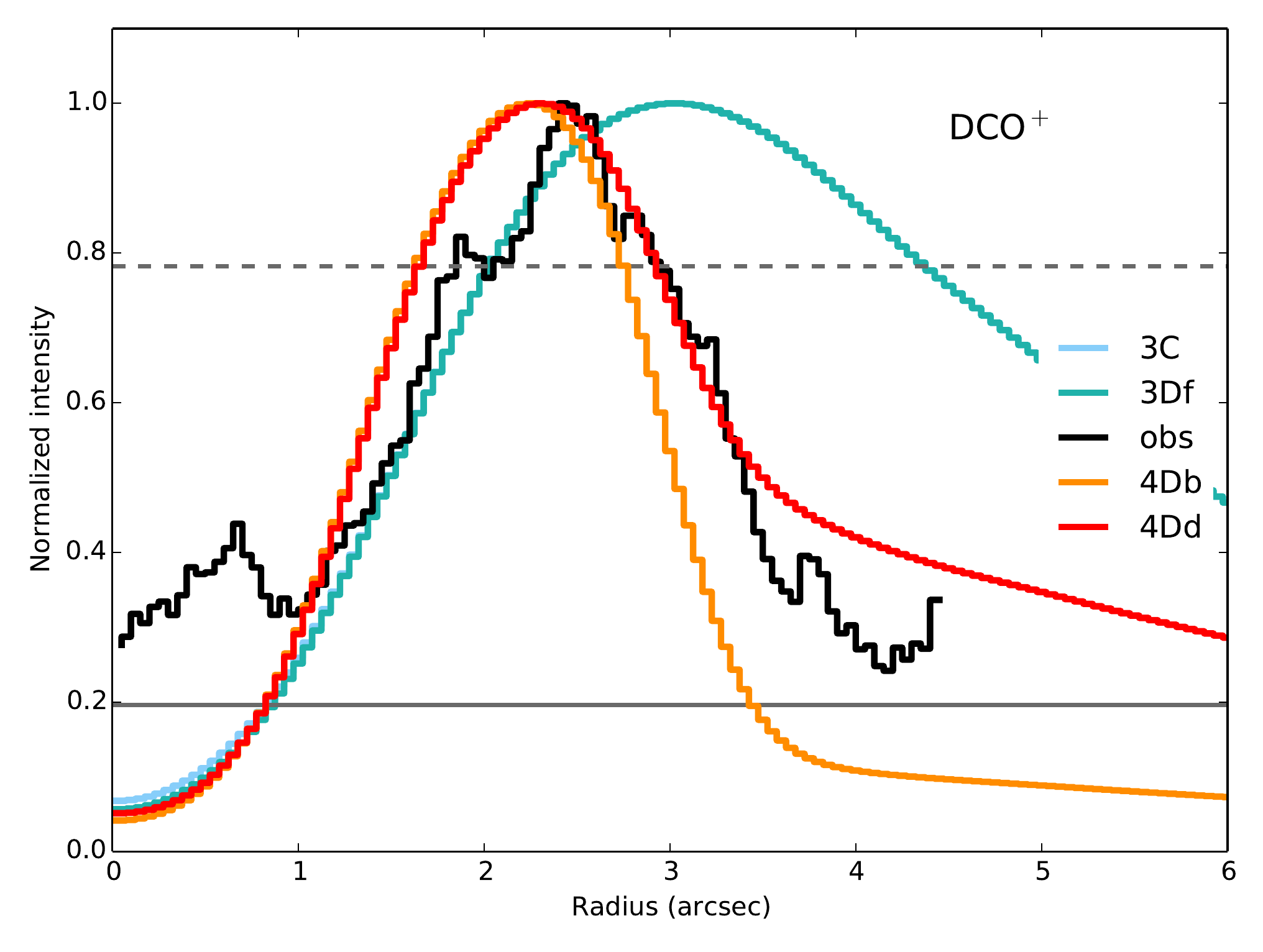}
\caption{Comparison between observations and radiative transfer modeling of the chemical network results. Here the best approximations are highlighted, with model 4Dd being the preferred model. The models are referred to by test, abundance profile (Constant ``C" or Drop ``D") and case. Thus, 4Dd is the model for test 4 with Drop abundance and case d. Black line shows the \ce{DCO+} observations integrated over the southern clump. Gray solid and dashed lines show the 1$\sigma$ and 4$\sigma$ levels, respectively.}
\label{figdcorad}
\end{figure}

\begin{figure}
\includegraphics[width=\columnwidth]{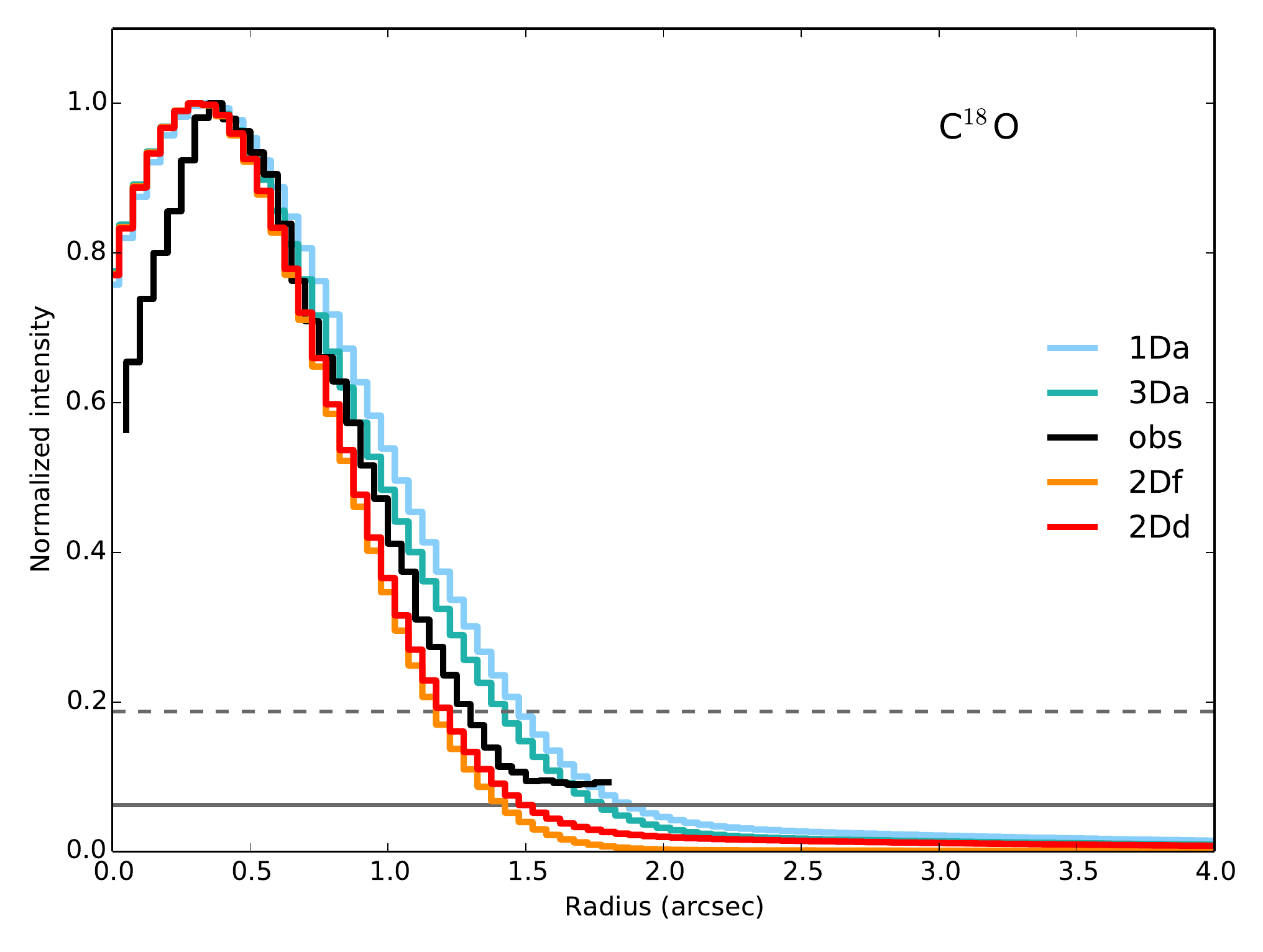}
\caption{Same as in Fig.~\ref{figdcorad} but for \ce{C^{18}O}, with 2Dd being the preferred model. Gray solid and dashed lines show the 1$\sigma$ and 3$\sigma$ levels, respectively. 2Dd, increased density profile, provides the best approximation to the observed \ce{C^{18}O}.}
\label{figc18orad}
\end{figure}

\subsubsection{Comparison with full chemical network}
\label{subsecmodcaveat}
To explore the limitations of the simple network, we compare the results to a full time-dependent deuterated chemical network, based on the RATE06 version of the UMIST Database for Astrochemistry \citep{woodall2007} extended with deuterium fractionation reactions \citep{mcelroy2013}. Models with only gas phase chemistry and including gas-grain balance (freeze-out, thermal desorption, and cosmic-ray-induced photodesorption) are run. The models start with the same initial abundandances (\ce{H2}:\ce{HD}:\ce{CO} = 1:3$\times$10$^{-5}$:1$\times$10$^{-4}$) and the same physical structure as the simple network. We find that the large network confirms both the general trend found in the simple network and also reproduces the peak position and abundance of \ce{DCO+} at the observed position for abundances extracted at early times ($\sim$10$^5$ yr).

Inspection of the full network shows that in addition to the reactions listed in Table~\ref{tabnet}, the \ce{HCO+ + D -> H + DCO+} reaction can become important if D/HD is large. The back reaction has a reaction barrier of $\sim$800 K, thus leading to fractionation of \ce{DCO+} \citep{adams1985}. In practice, this reaction is only relevant for
much lower densities than encountered in the VLA1623A envelope model. Similarly, dissociative recombination of \ce{H3+} with electrons cannot be neglected at low densities. At very early times ($< 10^3$ yr, depending on density), reactions with \ce{H+} and \ce{D+} become more significant than those with \ce{H3+} and \ce{H2D+}. Finally, in full
gas-grain models the bulk of the \ce{CO} is frozen out in the cold outer envelope resulting in very low \ce{DCO+} concentrations, unless an efficient non-thermal \ce{CO} desorption process is included.

\subsection{Emission profiles}
\label{subsecemission}
Aiming to better constrain the physical conditions of the disk-envelope interface we compare the models and observations of both \ce{DCO+} and \ce{C^{18}O} emission computed with RATRAN. For the comparison, the observed southern red-shifted clump in the ALMA 12-m array data for both lines is selected. The reason is two-fold: i) the red-shifted emission of the \ce{C^{18}O} disk suffers less absorption from the outer envelope than the blue-shifted lobe as noted above \citep{murillo2013b}; ii) the southern red-shifted clump traced by \ce{DCO+} is the strongest and most prominent. Similarly, for the model of \ce{C^{18}O} only one lobe of the disk is selected for comparison.

Consistent with the abundance plots, the constant \ce{CO} abundance profiles do not fit the observed \ce{DCO+} emission in any of the four tests, producing very broad peaks between 3$\arcsec$ and 5$\arcsec$ away from the source for tests 1, 2 and 3; and a peak at 1$\arcsec$ for test 4. In a similar manner, for the drop \ce{CO} abundance scenario, tests 1, 2 and 3 all produce peaks beyond 3$\arcsec$ with varying broadness, and thus do not approximate the observed \ce{DCO+} emission. This occurs for all examined cases (Fig.~\ref{figapdcorad}). Models 4Db and 4Dd approximate the observed \ce{DCO+} emission well (Fig.~\ref{figdcorad}, orange and red lines, respectively). These results are in agreement with the conclusions drawn from the concentrations in the analytics chemical network model (See Sec.~\ref{subsecmodphys}).

Comparing the results of the model with the observed \ce{C^{18}O} emission (Fig.~\ref{figc18orad}), we find that in all four tests a constant \ce{CO} abundance over-predicts the extent of \ce{C^{18}O}, with the emission well above our 3$\sigma$ level extending out to 3$\arcsec$ or further, whereas we observe it only out to less than 2$\arcsec$ from the central source. When the drop \ce{CO} abundance is introduced, the source profile of VLA1623 from \cite{jorgensen2002} overestimates the \ce{C^{18}O} emission in all cases, and generates a second peak at about 3$\arcsec$ at half the intensity of the central peak in almost all cases. The width of the drop does not have any significant effect on the modeled emission. The decreased density and decreased temperature profile tests do not fare better, as they again largely overestimate the \ce{C^{18}O} beyond 2$\arcsec$ and even produce secondary peaks. Case $a$ ($X_{\rm D}$ = 10$^{-6}$) for the original profile and decreased density seem to show some promise, however the modeled emission is around 3$\sigma$ at 1.5$\arcsec$ whereas observations at that radius are closer to 1$\sigma$. Surprisingly, we find that the increased density profile produces the best results with case $d$ ($X_{\rm D}$ = 10$^{-7}$), the same as that for \ce{DCO+} (Fig.~\ref{figc18orad}, red line). 

Hence we find that the \ce{DCO+} and \ce{C^{18}O} observed emission are not reproduced by the same physical structure (Figs.~\ref{figdcorad} and ~\ref{figc18orad}). \ce{DCO+} is well modeled by a temperature profile a factor of 1.5 lower than that needed to model \ce{C^{18}O}, whereas \ce{C^{18}O} is reproduced by a density profile an order of magnitude higher than required for \ce{DCO+}. The abundance in the drop $X_{\rm D}$ necessary to reproduce the emission is the same for both lines, $X_{\rm D}$=10$^{-7}$.

\begin{figure}
\includegraphics[width=\columnwidth]{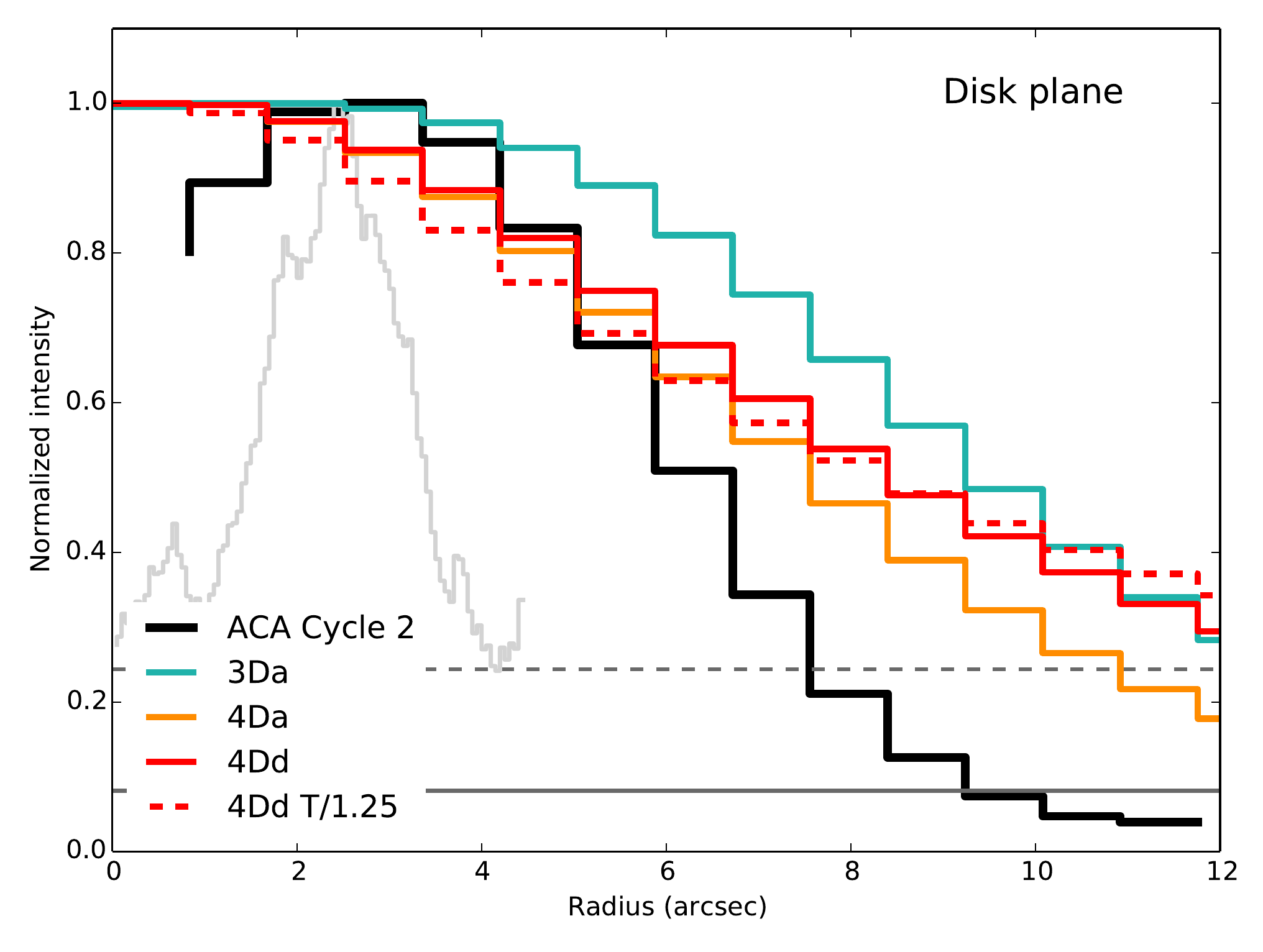}
\includegraphics[width=\columnwidth]{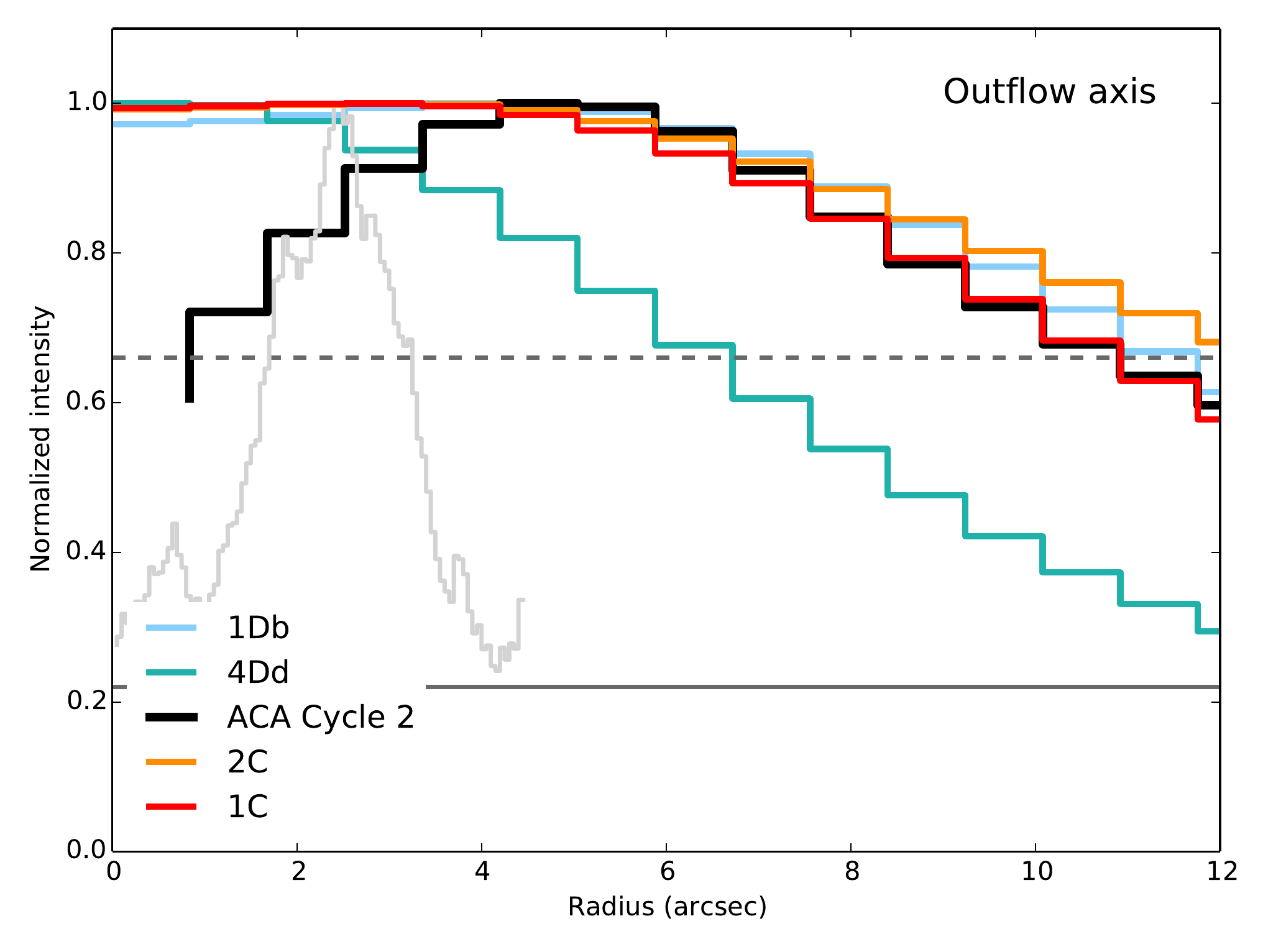}
\caption{Radial profiles of the \ce{DCO+} ACA observations along the disk plane (\textit{top}) and along the outflow axis perpendicular to the disk (\textit{bottom}) as shown in Fig.~\ref{figacamom}. Radial profile of the \ce{DCO+} Cycle 0 observations is shown in light gray. Observed radial profiles are overlaid with models convolved to the resolution of the ACA observations. Gray solid and dashed lines show the 1$\sigma$ and 3$\sigma$ levels, respectively, of the ACA emission. These results evidence that the \ce{DCO+} emission along the disk plane is best approximated by model 4Dd, whereas along the outflow is best described by model 1C.}
\label{figdcoradaca}
\end{figure}

\subsection{Large scale emission vs. model}
We also compare the radial integrated intensity profile of the \ce{DCO+} ACA observations with the model, both along the plane of the disk and perpendicular to the disk in the outflow direction, since these data probe larger spatial scales than the 12-m array data. For this purpose, the models were run through RATRAN's ray tracer \textit{sky} again but instead using the cell size of the ACA observations (0.84$\arcsec$) and then were convolved to match the resolution of the ACA observations. The observed radial profiles are extracted from the red-shifted emission.

The profile along the plane of the disk (peak at $\sim$4$\arcsec$) is well described by the same model as in the small scale, the drop \ce{CO} abundance profile for case d, with the temperature decreased by a factor of 1.5 (4Dd, Fig.~\ref{figdcoradaca}). The temperature decreased by a factor $<$1.5 does not provide a better estimate. The constant \ce{CO} abundance again overestimates the amount of \ce{DCO+} produced in the outer regions.

On the other hand, the integrated intensity profile along the outflow direction (peak at distances $\approx$5$\arcsec$) is very well described by the model 1C (Fig.~\ref{figdcoradaca}). This indicates that in regions that are not shadowed by the disk, the production of \ce{DCO+} is as expected from dust continuum models.

\begin{figure}
\includegraphics{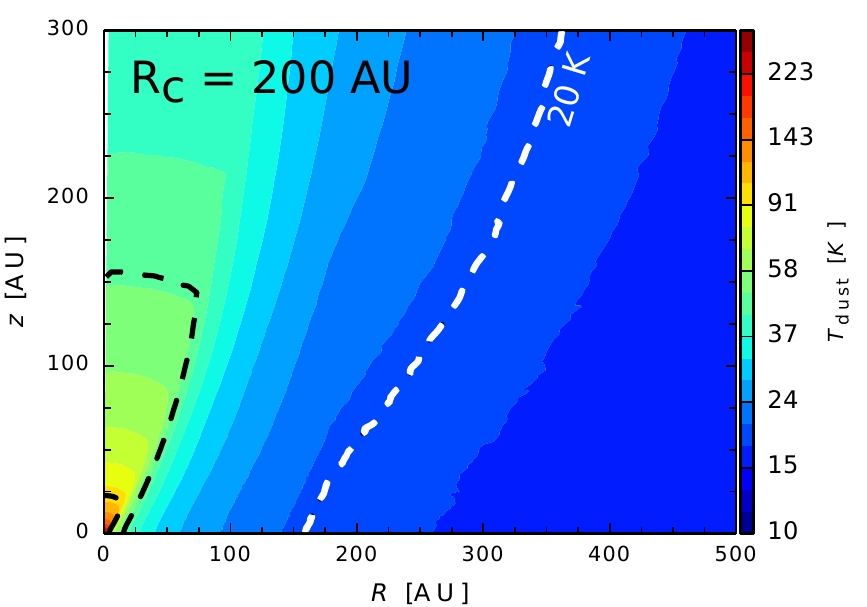}
\includegraphics{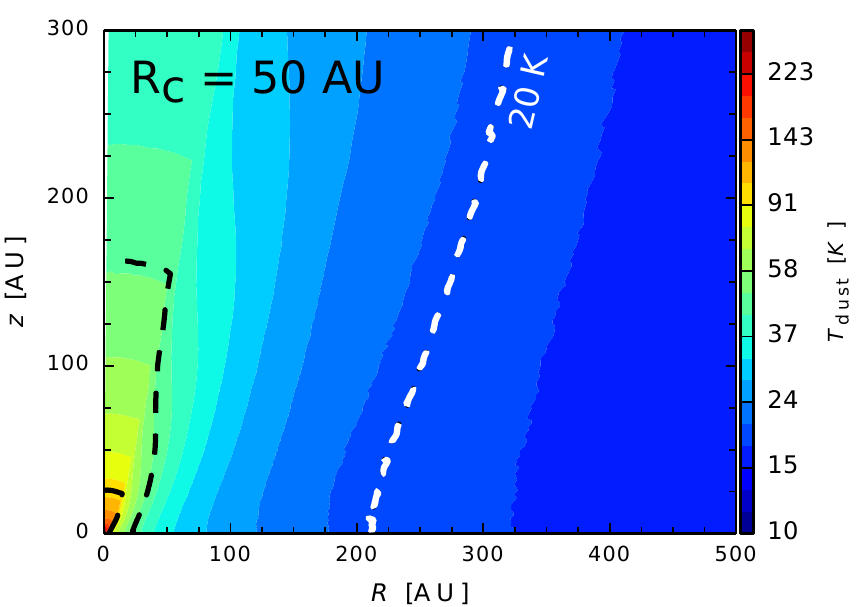}
\includegraphics{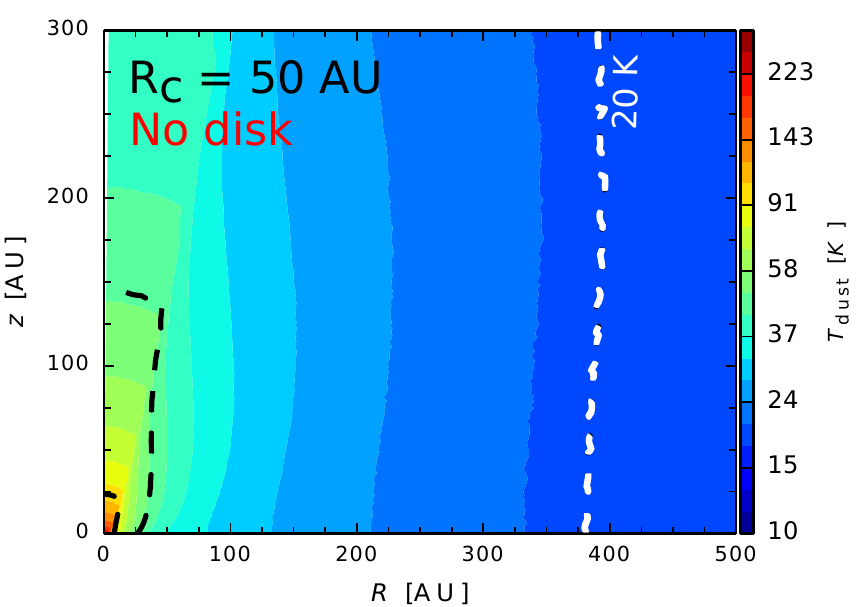}
\caption{Radiative 2D models based on \cite{harsono2013}. For all models an envelope with $M_{\rm env}$ = 1 M$_{\odot}$ irradiated by a 1 $L_{\odot}$ protostar is assumed, having an outflow cavity with a full opening angle of 30$^{\circ}$. The dashed lines indicate the temperature contours for 100, 50 and 20 K. The top and middle panels include a 180 AU and 0.02 M$_{\odot}$ disk assuming a 0.1 AU scale height, with centrifugal radius $R_{\rm c}$ = 200 AU (\textit{top}) and 50 AU (\textit{center}) orientated in the horizontal direction. The outflow cavity is oriented in the vertical direction. The bottom panel shows the envelope without a disk.}
\label{figmoddiskenv}
\end{figure}

\section{Discussion}
\label{secdis}
The results of our modeling show that the position of the \ce{DCO+} emission ($\sim$2.5$\arcsec$ = $\sim$300 AU) is closer to VLA1623A than expected based on the spherically symmetric dust continuum radiative transfer modeling from DUSTY ($\sim$4$\arcsec$ = $\sim$480 AU). The results further show that the observed emission peaks at a dust temperature range of 11-16 K. Comparison of the radiative transfer results from RATRAN to the \ce{C^{18}O} emission from the disk suggest that the disk has a density higher by one order of magnitude than the emitting structure traced by \ce{DCO+}.

A possible explanation would be that VLA1623A is a Very Low Luminosity Object (VeLLO, \citealt{young2004}, \citealt{dunham2008}) undergoing episodic accretion and just coming out of the quiescent phase. This is highly unlikely, however, since \cite{johnstone2013} find that the timescale for dust and gas to heat up after an accretion burst is short, on the order of hours to weeks. VLA1623A has been reported of having a bolometric luminosity between 0.4-2 L$_{\odot}$ from early observations to more recent work \citep{andre1993, froebrich2005, murillo2013a, chen2013}, in contrast to the expected luminosity of 10$^{-2}$ L$_{\odot}$ for VeLLOs. Hence VLA1623A has not recently come out of the quiescent accretion phase and the location of the observed \ce{DCO+} emission can not be attributed to the relic of a previous phase of decreased accretion.

A more plausible explanation to the position and temperature of the region containing the \ce{DCO+} emission is disk shielding. \ce{DCO+} is observed to border VLA1623A's \ce{C^{18}O} disk in our present data. This could shield the outer parts of the disk from heating by the central protostar, thus moving in the freeze-out zone of \ce{CO}, enhancing the production of \ce{DCO+} closer to the protostar. This scenario is further supported by the result of our simple chemical model that the \ce{C^{18}O} disk is more dense than the region of the \ce{DCO+} emission and than expected from \cite{jorgensen2002}'s envelope density profile of VLA1623.

We test the possibility of the disk-shielding scenario with radiative 2D disk plus envelope models using radiative transfer methods as in \cite{harsono2013} (Fig.~\ref{figmoddiskenv}). For the models, we assume a central protostellar source of 1 L$_{\odot}$ \citep{murillo2013a, chen2013}, an envelope with a mass of 1 M$_{\odot}$ \citep{andre1993, froebrich2005}, with the addition of a disk mass of 0.02 M$_{\odot}$ and radius of 180 AU \citep{murillo2013b}. The outflow cavity is assumed to have an opening angle of 30$^{\circ}$. A thin disk with a scale height of 0.1 AU is adopted. Disk flaring is not included since we have no information on the flaring of VLA1623A's disk and the thin disk model approximates the \ce{C^{18}O} kinematics well. Two values for the centrifugal radius $R_{\rm c}$, 200 AU and 50 AU, are chosen. Within the centrifugal radius the velocity structure of the disk is Keplerian. We find that even for such a thin disk the temperature along the plane of the disk is lower than for the envelope at the same radius (Fig.~\ref{figmoddiskenv}, top and middle panels), thus moving the \ce{CO} freeze-out zone closer to the protostar along the edge of the disk than in other regions of the core. For either centrifugal radius the temperature beyond 200 AU drops well below 20 K. Finally, we test whether the presence of a disk makes a difference in the location of the \ce{CO} freeze out region. Figure~\ref{figmoddiskenv} bottom panel, shows the model with the same conditions as in Figure~\ref{figmoddiskenv} middle panel, but without the disk. This shows that omitting the disk causes the \ce{CO} freeze out region to move outward to $\sim$400 AU, 150 AU further out than the models including a disk. 

The temperature conditions required to reproduce the observed \ce{DCO+} emission with our simple chemical model are therefore in agreement with the results obtained from the 2D radiative transfer disk plus envelope model. These results strongly support the scenario where a disk can shield the regions at its edge from heating by the protostar. This shielding causes the \ce{CO} freeze-out region to move inward toward the edge of the disk, bringing low-temperature enhanced molecules, such as \ce{DCO+}, closer to the central protostar along the plane of the disk. However, the rest of the envelope is largely unaffected by the disk and a shell of molecules such as \ce{DCO+} forms at a radius predicted by spherically symmetric models. The ACA observations of \ce{DCO+} toward VLA1623A provide further evidence for this effect. These results show that the \ce{DCO+} emission peaks closer to the source along the disk plane than along the outflow direction.

\begin{figure}
\includegraphics[width=\columnwidth]{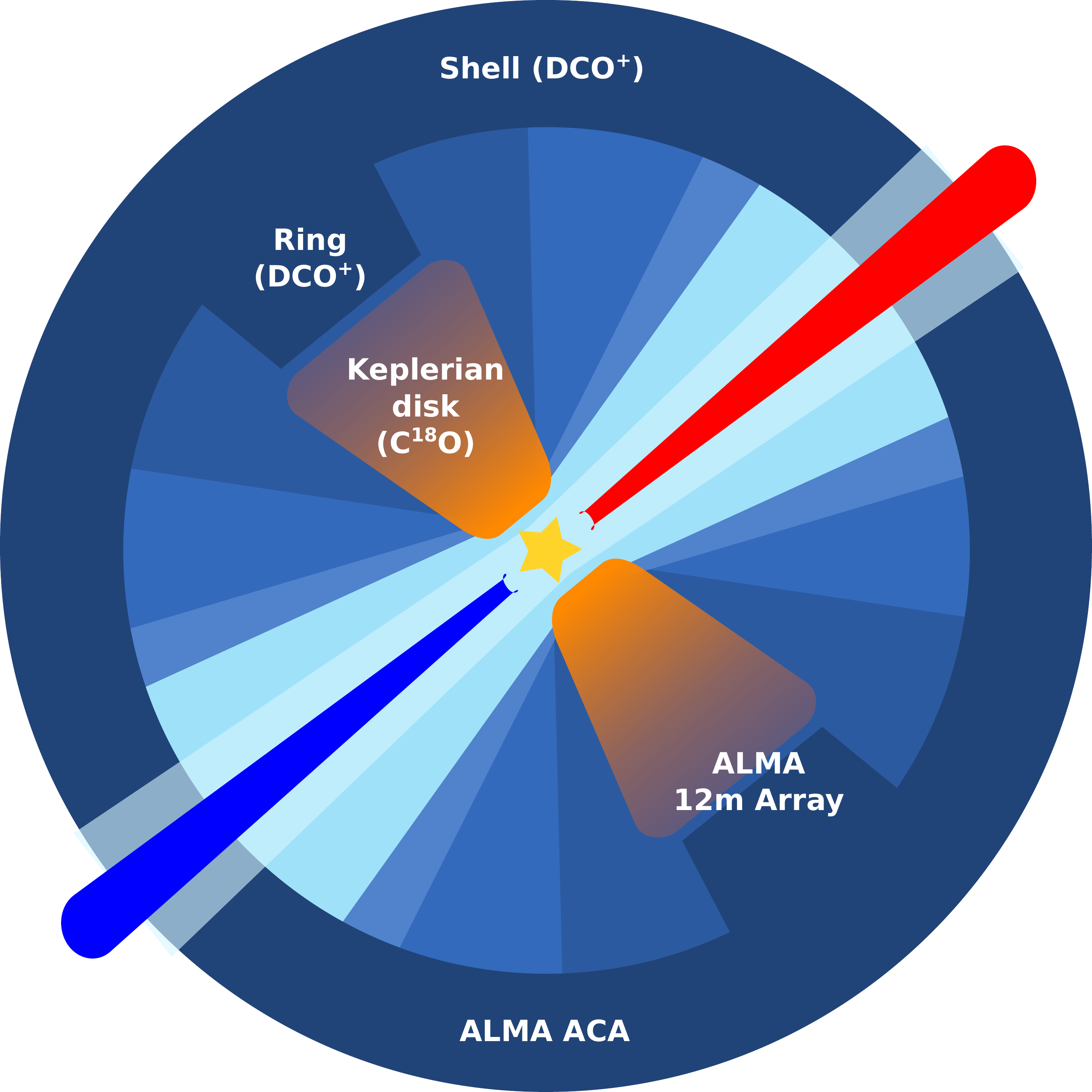}
\caption{Cartoon of the disk and envelope of VLA1623A. The ring and disk, traced in \ce{DCO+} and \ce{C^{18}O} respectively, are observed with ALMA 12-m array, while the outer spherically symmetric shell, traced in \ce{DCO+}, is observed with ALMA ACA.}
\label{figcartoon}
\end{figure}

\section{Conclusions}
\label{secconc}
This work presents the results and analysis of ALMA Cycle 0 Early Science Band 6 observations of \ce{DCO+} (3-2) in the extended configuration toward VLA1623A probing subarcsec scales, as well as Cycle 2 ACA data probing the larger scales. A simple chemical network was setup taking ortho- and para-\ce{H2} into account in the rate-determining reactions. The density and temperature profile of VLA1623A was obtained from fitting the SED and dust continuum data with radiative transfer modeling using DUSTY \citep{jorgensen2002}. Our simple chemical model coupled with VLA1623A's physical structure served as input for line radiative transfer calculations with RATRAN with 2-D ray-tracing. The \ce{CO} abundance, density and temperature profiles were altered to study the effect of each parameter on the location of the observed \ce{DCO+} peak. The results of our observations and analysis can be thus summarized:

1. \ce{DCO+} is observed to border the \ce{C^{18}O} disk around VLA1623A. Both emission lines show similar velocity gradients (blue-shifted to the north and red-shifted to the south), with \ce{DCO+} emission between 2.8 to 5.2 km~s$^{-1}$. The PV diagrams of \ce{C^{18}O} and \ce{DCO+} suggest that both emission lines are well described by Keplerian rotation. However, the \ce{DCO+} emission is weak, thus no further kinematical analysis was carried out, and whether the disk extends out to 300 AU cannot be confirmed.

2. Using a simple chemical network with the inclusion of ortho- and para-\ce{H2} as well as non-LTE line radiative transfer, we model the observed \ce{DCO+} emission. We find that using a constant \ce{CO} abundance predicts a \ce{DCO+} peak at around 4$\arcsec$, twice further out than observed, irrespective of the adopted o/p ratio and density profile. Our model results show that a drop \ce{CO} abundance with a decreased temperature profile by a factor of 1.5 generates a peak at the same position as the observed emission. Thus, the observed \ce{DCO+} peak is closer to VLA1623A and with a lower temperature (11-16 K) than that expected from a spherically symmetric physical structure constrained by continuum data and source SED.

3. The observed \ce{DCO+} and \ce{C^{18}O} emission are not described by the same physical structure. In our model, the \ce{C^{18}O} emission is well reproduced by a drop \ce{CO} abundance with the density profile increased by one order of magnitude at 1$\arcsec$ ($\leq$120 AU) radii, in comparison with the density profile needed to reproduce the observed \ce{DCO+} emission. A constant \ce{CO} abundance and a decreased temperature profile over-predict the extent of the \ce{C^{18}O} emission.

4. Disk-shielding is the best possible explanation for the observed \ce{DCO+} emission toward VLA1623A. Disk-shielding causes the inward shift of the \ce{CO} freeze-out region along the plane of the disk, lowering the dust temperature to $<$20 K, generating a ring of molecules whose abundance is enhanced by low temperatures such as \ce{DCO+}. The rest of the envelope is largely unaffected by the disk, thus the \ce{CO} freeze-out shell predicted by spherically symmetric radiative models is expected to be located further out. This prediction is confirmed by our recent ALMA Cycle 2 ACA observations, which show that the \ce{DCO+} emission along the outflow axis lies at larger radii, $\sim$5$\arcsec$, consistent with constant \ce{CO} abundance models without any alteration to \cite{jorgensen2002}'s temperature and density profile of VLA1623.

5. The disk-envelope interface in VLA1623A is shown to have a broken transition in density and temperature, with the impact generated by the presence of a disk being observable from small to large scales.

Our observations and modeling results for VLA1623A show the disk-envelope interface to have different physical conditions than other parts of the envelope. Our results also highlight the drastic impact that the disk has on the temperature structure at $\sim$100 AU along the plane of the disk, with the effect being observable even at large scales. We suggest performing further observations to determine whether the unequal physical and chemical conditions observed in the disk-envelope interface of VLA1623A is a common phenomenon in protostellar systems with rotationally supported disks or a special condition of the present source.

\begin{acknowledgements}
This paper made use of the following ALMA data: ADS/JAO.ALMA 2011.0.00902.S and 2013.1.01004.S. ALMA is a partnership of ESO (representing its member states), NSF (USA), and NINS (Japan), together with NRC (Canada) and NSC and ASIAA (Taiwan), in cooperation with the Republic of Chile. The Joint ALMA Observatory is operated by ESO, AUI/NRAO, and NAOJ. The 2011.0.00902.S data was obtained by N.M.M. while she was a Master student at National Tsing Hua University, Taiwan, under the supervision of S.P.L. The authors thank J. C. Mottram for his help in using RATRAN; and M. Hogerheijde for permission to use the 2-D version of RATRAN. Astrochemistry in Leiden is supported by the Netherlands Research School for Astronomy (NOVA), by a Royal Netherlands Academy of Arts and Sciences (KNAW) professor prize, and by the European Union A-ERC grant 291141 CHEMPLAN. This work is supported by grant 639.041.335 from the Netherlands Organisation for Scientific Research (NWO) and by the Netherlands Research School for Astronomy (NOVA) and by the Space Research Organization Netherlands (SRON). S.P.L. acknowledges support from the Ministry of Science and Technology of Taiwan with Grants MOST 102-2119-M-007-004-MY3.
\end{acknowledgements}

\bibliographystyle{aa}
\bibliography{vla1623_dco_langed.bib}

\Online 

\begin{appendix}
\section{Modeling results: additional figures}
\label{apsecmod}
\begin{figure*}
\includegraphics[width=\textwidth]{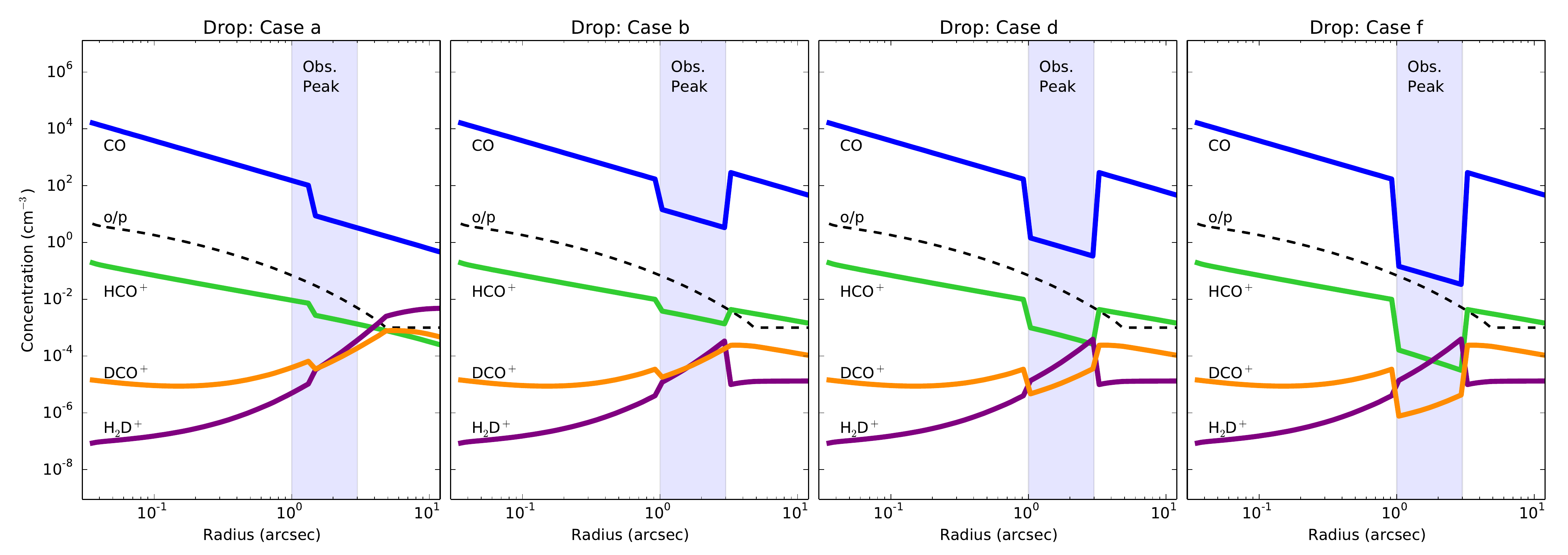}
\caption{Chemical network results for Drop \ce{CO} abundance with the Test 1 source profile from \cite{jorgensen2002} for four cases (Table~\ref{tabcase}). Case d is the same as in the top row center panel of Fig.~\ref{figresults}.}
\label{figdropcaset1}
\end{figure*}

\begin{figure*}
\includegraphics[width=\textwidth]{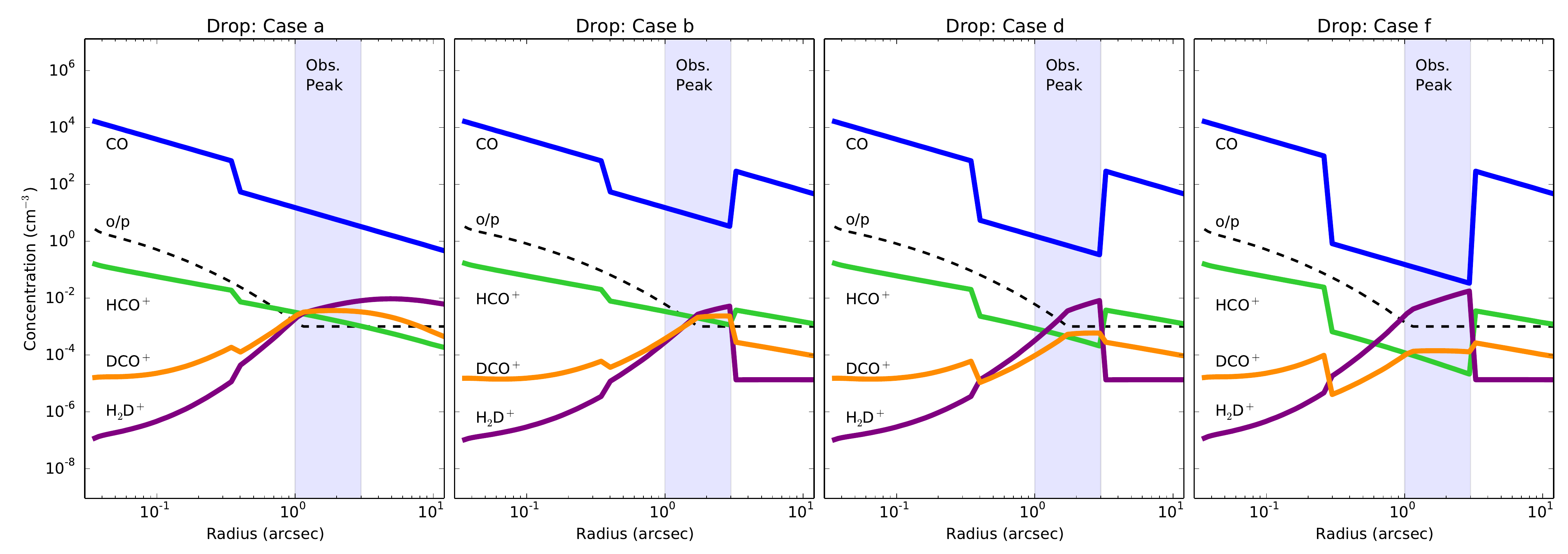}
\caption{Chemical network results for Drop \ce{CO} abundance with the decreased-temperature (Test 4) source profile for four cases (Table~\ref{tabcase}). Case d is the same as in the bottom row center panel of Fig.~\ref{figresults}.}
\label{figdropcase}
\end{figure*}

\begin{figure*}
\includegraphics[width=\textwidth]{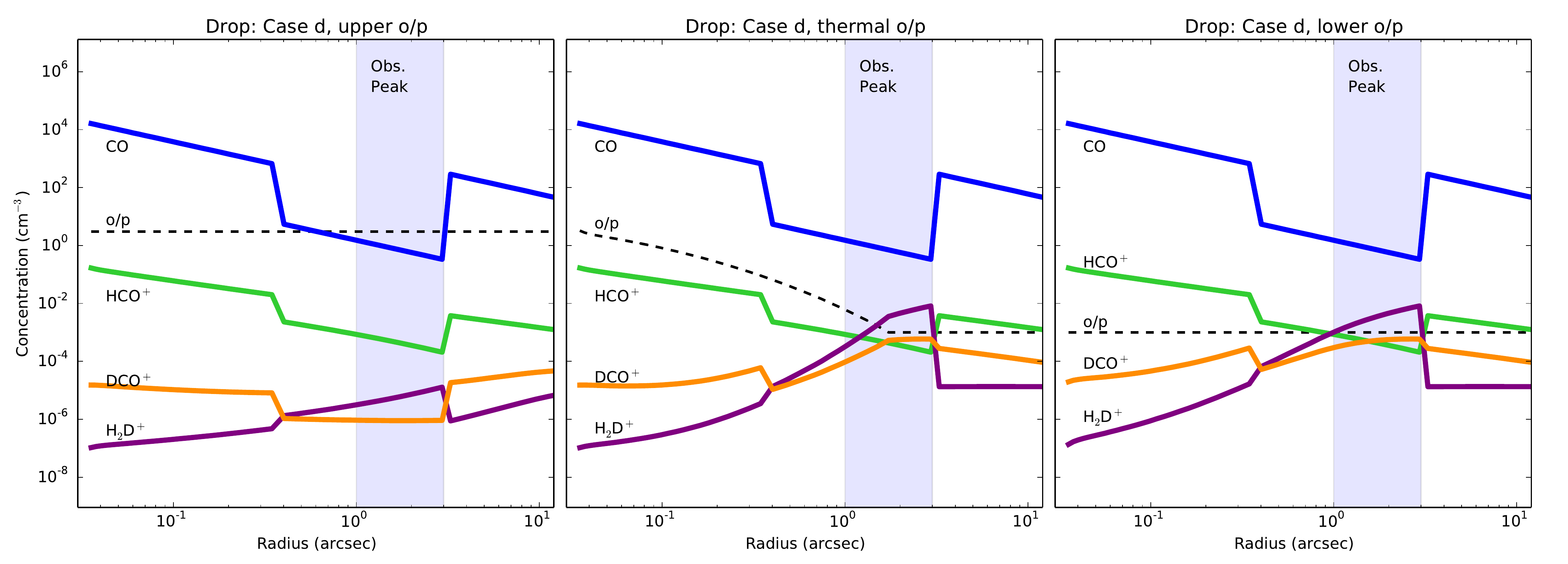}
\caption{Chemical network results for Drop \ce{CO} abundance with the decreased-temperature source profile (Test 4) for case a with upper-limit (\textit{left}), thermal (\textit{center}) and lower-limit (\textit{right}) o/p ratio.}
\label{figdropoplim}
\end{figure*}

\begin{figure*}
\includegraphics[width=\textwidth]{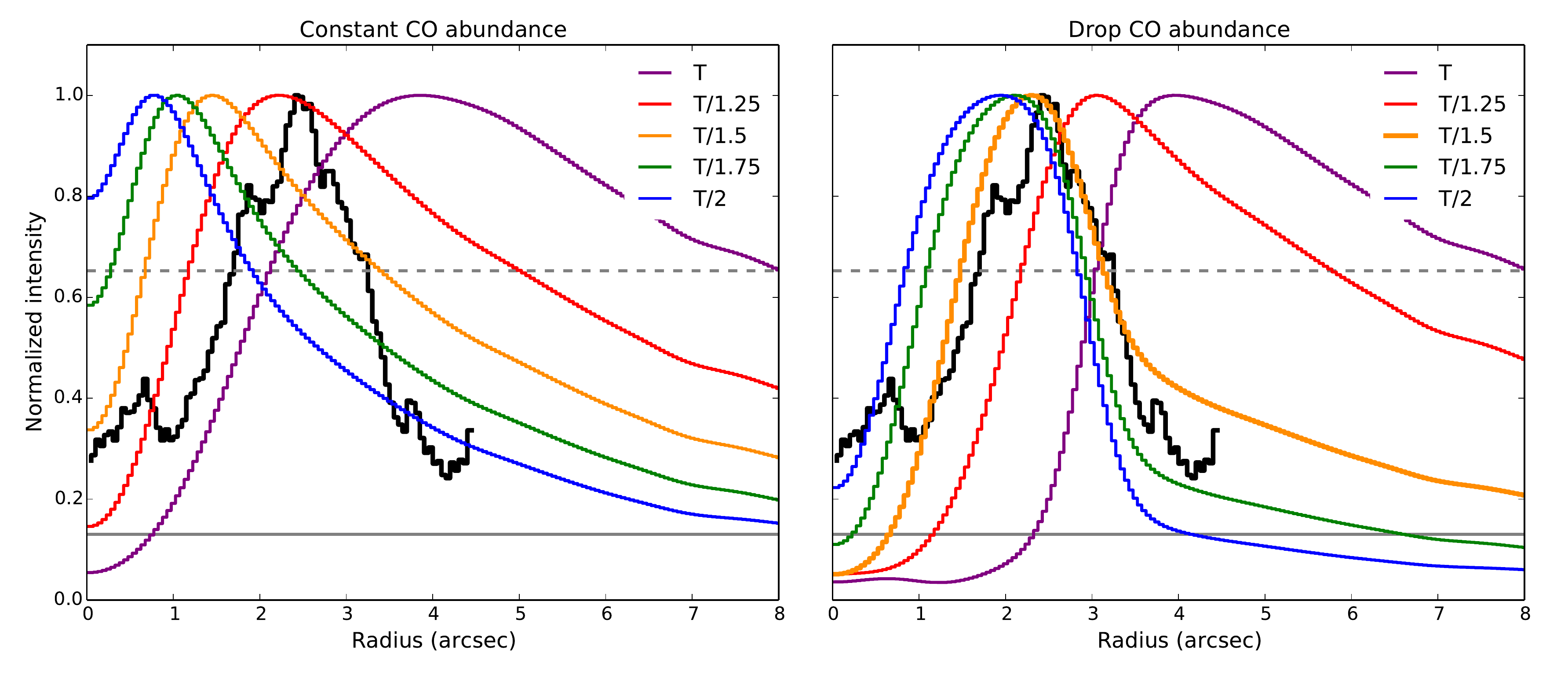}
\caption{\ce{DCO+} Cycle 0 observations compared with the radial profile of the constant and drop \ce{CO} abundance models, showing the effect of the factor used to decrease the temperature, for case d. A factor of 1.5 with drop \ce{CO} abundance provides the best approximation to the observed radial profile.}
\label{figaptemp}
\end{figure*}

\begin{figure*}
\includegraphics[width=\textwidth]{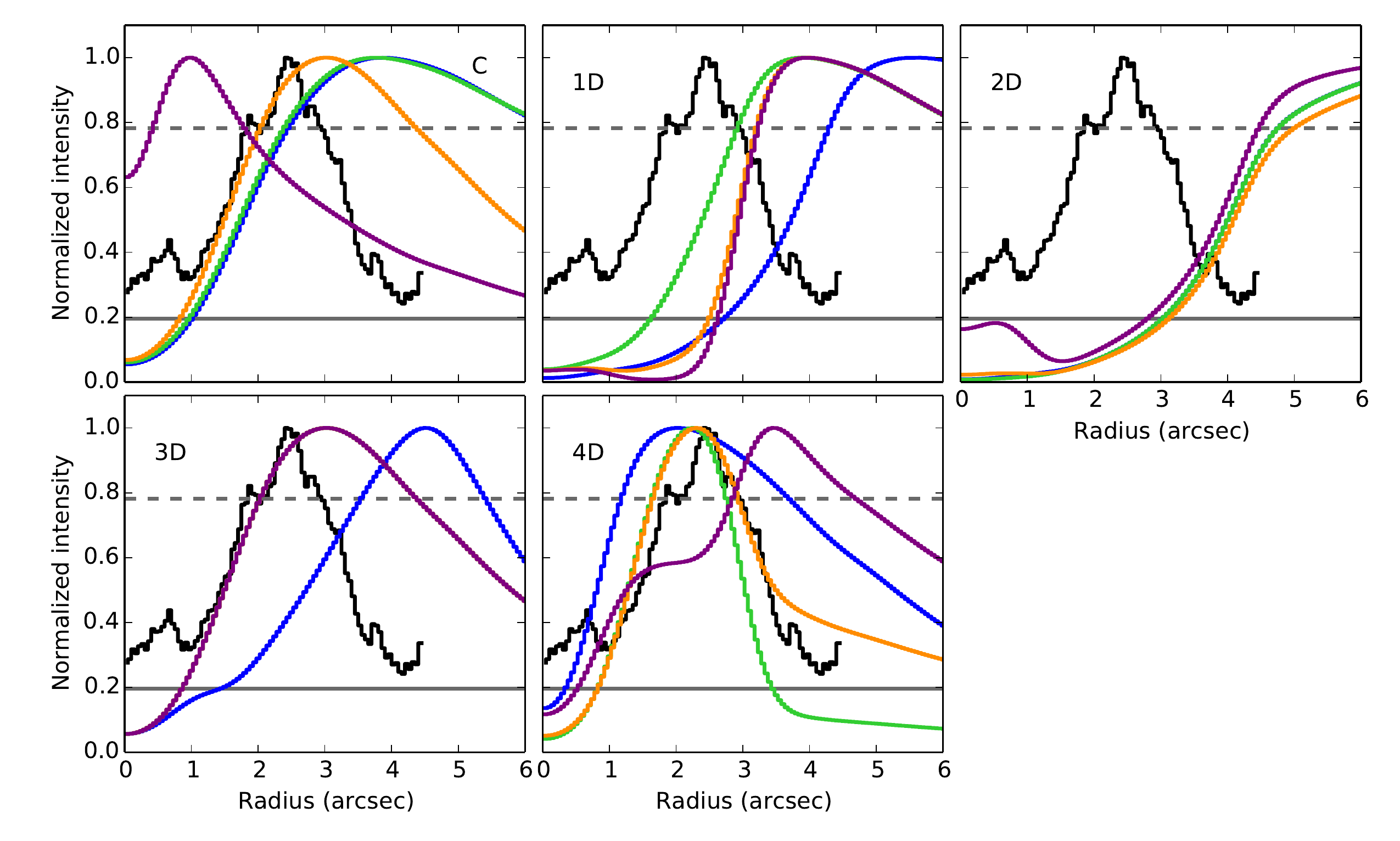}
\caption{Comparison of the observations and radiative transfer modeling of the chemical network results. The black line shows the \ce{DCO+} observations integrated over the southern clump, with the gray solid lines showing the 1$\sigma$ and 4$\sigma$ levels, respectively. Models are referred to by test, abundance profile (Constant "C" or Drop "D") and case. Color lines show tests 1 (blue), 2 (green), 3 (orange) and 4 (purple) for the top left panel, and cases a (blue), b (green), d(orange) and f (purple) for the other four panels.}
\label{figapdcorad}
\end{figure*}

\begin{figure*}
\includegraphics[width=\textwidth]{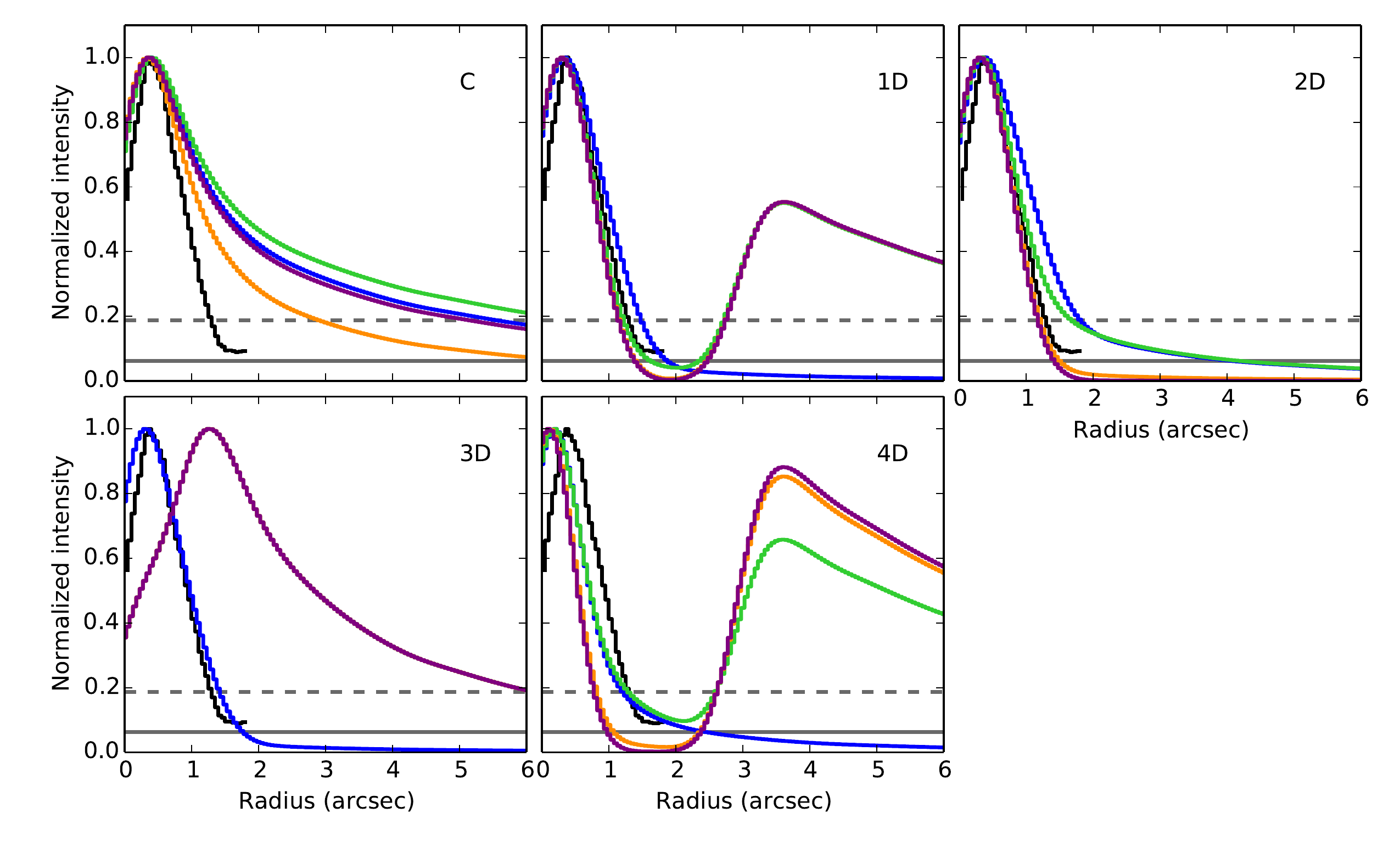}
\caption{As in Fig.~\ref{figapdcorad} but for \ce{C^{18}O}. Gray solid and dashed lines show the 1$\sigma$ and 3$\sigma$ levels respectively. Note that constant \ce{CO} abundance over-predicts the amount of \ce{C^{18}O} emission.}
\label{figapc18orad}
\end{figure*}

Here we present additional figures illustrating details discussed in Section~\ref{subsecmod} and ~\ref{subsecemission}. Figures~\ref{figdropcaset1} and~\ref{figdropcase} show the effect of \ce{CO} abundance $X_{\rm CO}$ in the drop on the \ce{DCO+} peak for Test 1 and 4, respectively. Figure~\ref{figdropoplim} highlights the effect of the upper and lower o/p limits on the \ce{DCO+} abundance for the best approximation models to the \ce{DCO+} observations. The effect of the factor used to decrease the temperature profile on the peak position of the modeled \ce{DCO+} is explored in Figure~\ref{figaptemp}. Figures~\ref{figapdcorad} and ~\ref{figapc18orad} give a sampling of the observations compared with the radiative transfer modeling of the chemical network results for the different tests, \ce{CO} abundance profiles and cases.

\end{appendix}

\end{document}